\documentclass[trackchanges]{aastex631}
\usepackage{amsmath}

\DeclareRobustCommand{\ion}[2]{\textup{#1\,\textsc{\lowercase{#2}}}}
\newcommand{\HI}{\ion{H}{i}}
\newcommand{\NHI}{N_{\scriptsize \ion{H}{i}}}
\newcommand{\NHIi}{N_{\scriptsize \ion{H}{i},i}}

\newcommand{\CHIi}{C_{\scriptsize \ion{H}{i},i}}
\newcommand{\CHIone}{C_{\scriptsize \ion{H}{i},1}}
\newcommand{\CHItwo}{C_{\scriptsize \ion{H}{i},2}}
\newcommand{\CHIthr}{C_{\scriptsize \ion{H}{i},3}}

\newcommand{\NHtot}{N_\mathrm{H}^\mathrm{tot}}

\newcommand{\NH}{N_\mathrm{H}}
\newcommand{\NHtwo}{N_\mathrm{H_{2}}}
\newcommand{\Htwo}{\mathrm{H}_{2}}

\newcommand{\WCO}{W_\mathrm{CO}}
\newcommand{\WHI}{W_{\scriptsize \ion{H}{i}}}

\newcommand{\WHItwothr}{W_{\scriptsize \ion{H}{i},2+3}}
\newcommand{\WHItwo}{W_{\scriptsize \ion{H}{i},2}}
\newcommand{\THI}{\tau_{\scriptsize \ion{H}{i}}}
\newcommand{\VHI}{V_{\scriptsize \ion{H}{i}}}
\newcommand{\HIs}{\mbox{{\scriptsize H}{\tiny\,I}}}
\newcommand{\HIalt}{\mbox{H\footnotesize\,I}} 

\shorttitle{Gas and CRs in MBM 53-55 and Pegasus loop Revealed}
\shortauthors{Mizuno et al.}
\graphicspath{{./}{figures/}}

\begin{document}

\title{Gas and Cosmic-Ray Properties in the MBM 53, 54, and 55 Molecular Clouds and the Pegasus Loop
as Revealed by $\HIalt$ Line Profiles, Dust, and Gamma-Ray Data}

\author[0000-0001-7263-0296]{T.~Mizuno}
\email{mizuno@astro.hiroshima-u.ac.jp}
\affiliation{Hiroshima Astrophysical Science Center, Hiroshima University, Higashi-Hiroshima, Hiroshima 739-8526, Japan}
\author{K.~Hayashi}
\affiliation{Institute of Space and Astronautical Science, Japan Aerospace Exploration Agency, 3-1-1 Yoshinodai, Chuo-ku, Sagamihara, Kanagawa 252-5210, Japan}
\author{J.~Metzger}
\affiliation{Department of Physics, The University of Chicago, Chicago, Illinois 60637, USA}
\author{I.~V.~Moskalenko}
\affiliation{W. W. Hansen Experimental Physics Laboratory, Kavli Institute for Particle Astrophysics and Cosmology, Stanford University, Stanford, CA 94305, USA}
\author{E.~Orlando}
\affiliation{W. W. Hansen Experimental Physics Laboratory, Kavli Institute for Particle Astrophysics and Cosmology, Stanford University, Stanford, CA 94305, USA}
\affiliation{Department of Physics, University of Trieste and INFN, I-34127 Trieste, Italy}
\affiliation{Eureka Scientific, Oakland, CA 94602-3017, USA}
\author{A.~W.~Strong}
\affiliation{Max-Planck Institut f\"ur extraterrestrische Physik, D-85748 Garching, Germany}
\author{H.~Yamamoto}
\affiliation{Department of Physics and Astrophysics, Nagoya University, Chikusa-ku Nagoya 464-8602, Japan}



\begin{abstract}
In studying the interstellar medium (ISM) and Galactic cosmic rays (CRs),
uncertainty of the interstellar gas density has always been an issue. 
To overcome this difficulty, 
we used a component decomposition of the 21-cm $\HI$ line emission and 
used the resulting gas maps in an analysis of $\gamma$-ray data obtained by the 
\textit{Fermi} Large Area Telescope (LAT) for the
MBM~53, 54, and 55 molecular clouds and the Pegasus loop.
We decomposed the ISM gas into intermediate-velocity clouds, narrow-line and optically thick $\HI$,
broad-line and optically thin $\HI$, CO-bright $\Htwo$, and CO-dark $\Htwo$
using detailed correlations with the
$\HI$ line profiles from the HI4PI survey, the \textit{Planck} dust-emission model, and 
the \textit{Fermi}-LAT $\gamma$-ray data.
We found the fractions of 
optical depth correction to the $\HI$ column density
and CO-dark $\Htwo$ to be nearly equal.
We fitted the CR spectra directly measured at/near the Earth and the measured $\gamma$-ray emissivity spectrum
simultaneously.
We obtained a spectral break in the interstellar proton spectrum at ${\sim}$7~GeV, and
found the $\gamma$-ray emissivity normalization agrees with the AMS-02 proton spectrum within 10\%,
relaxing the tension with the CR spectra previously claimed.
\end{abstract}



\section{Introduction}
Interstellar space in the Milky Way is permeated with ordinary matter (gas or dust), which is known as the interstellar medium (ISM).
It also contains high-energy charged particles known as cosmic rays (CRs),
an interstellar radiation field (ISRF), and a magnetic field.
These components are mutually interacting, and they play important roles in many physical and chemical processes (e.g., star formation).
Hence, they have been studied in various wavebands from radio to $\gamma$-rays
\citep[for a review, see, e.g.,][]{Ferriere2001}.

Cosmic $\gamma$-ray emission (with photon energies $E \gtrsim 100~\mathrm{MeV}$)
is known to be a powerful probe for studying the ISM and Galactic CRs.
High-energy CR protons and electrons interact with the interstellar gas 
or the ISRF
and produce $\gamma$-rays through nucleon--nucleon interactions, electron bremsstrahlung, 
and inverse-Compton (IC) scattering.
Because the ISM is essentially transparent to those $\gamma$-rays
\citep[e.g.,][]{Moskalenko2006},
we can study the ISM distribution via $\gamma$-ray observations.
Because the $\gamma$-ray production cross-section is independent of the chemical or 
thermodynamic state of the interstellar gas, cosmic $\gamma$-rays have been recognized 
as a unique tracer of the total column density of gas, regardless of its atomic or molecular state.
If observations in other wavebands can be used to estimate the gas column density accurately,
the CR spectrum and intensity distribution can be examined as well. 

Usually, the distribution of neutral atomic hydrogen ($\HI$) is measured directly via 21-cm line 
surveys \citep[e.g.,][]{Dickey1990, Kalberla2009},
assuming the optically thin approximation,
and the distribution of molecular hydrogen ($\Htwo$) is 
estimated indirectly from carbon monoxide (CO) line-emission surveys \citep[e.g.,][]{Dame2001},
assuming a linear conversion factor.
However, some fraction of the ISM gas in optically thick $\HI$ or CO-dark $\Htwo$ phases
may be missed by these line surveys.
Such ``dark gas" can be studied
using total gas tracers such as dust extinction, reddening, and emission \citep[e.g.,][]{Reach1994}
and $\gamma$-rays \citep[e.g.,][]{Grenier2005}.
The work by \citet{Grenier2005}
has been confirmed and improved by subsequent observations
with the \textit{Fermi} Large Area Telescope \citep[LAT;][]{Atwood2009}.
In addition, the \textit{Planck} mission has provided an
all-sky model of thermal emission from dust \citep{Planck2011,Planck2014}
that is useful for the study of the ISM gas distribution 
because of its sensitivity and high angular resolution. 

Despite these efforts, uncertainties in the ISM gas column density and the CR intensity are still uncomfortably large,
by as much as a factor of {$\sim$}50\% even in local environment \citep[see, e.g.,][]{Grenier2015}.
This is mainly due to the uncertainty in the spin temperature ($T_\mathrm{s}$) of the $\HI$ gas,
which affects the conversion from the 21-cm line intensity to the $\HI$ gas column density.
To cope with this difficulty, some authors \citep[e.g.,][]{Mizuno2016, Hayashi2019} 
have proposed to treat areas with high dust temperatures as optically thin $\HI$ and 
have used this assumption in analyzing \textit{Fermi}-LAT $\gamma$-ray data.
However, their method cannot distinguish gas phases along the line of sight,
and hence, it is not applicable to the Galactic plane. Also, the composition of the dark gas
(i.e., the fractions of the optically thick $\HI$ and CO-dark $\Htwo$) 
is quite uncertain and is controversial.
For example, while \citet{Fukui2015} proposed that optically thick $\HI$ dominates dark gas,
\citet{Murray2018} claimed that dark gas 
is mainly molecular.
Again, this is because the value of $T_\mathrm{s}$ is usually unknown, and 
neither dust nor $\gamma$-rays can distinguish between atomic and molecular gas phases.

$\HI$ absorption features (optical depth profile) are often well represented by Gaussians,
supporting that gas motions within $\HI$ clouds have a random velocity distribution.
$\HI$ emission profiles of most sources can also be decomposed into Gaussians, and 
components with narrow or broad linewidths 
could be assumed to mainly arise 
from the cold neutral medium (CNM) or warm neutral medium (WNM), respectively \citep[e.g.,][]{Kalberla2020}.
Recently, \citet{Kalberla2018} analyzed the all-sky HI4PI survey data \citep{HI4PI} and decomposed the $\HI$ 21-cm line emission into
Gaussian lines by taking account of spatial coherence. 
Although their study uses emission spectra only and hence suffers from 
systematic uncertainties in separating the CNM and WNM,
it allows them to study $\HI$ line profies over the entire sky.
Subsequently, \citet{Kalberla2020} found that narrow-line $\HI$ gas is associated with
the dark gas estimated from infrared dust-emission maps by \citet{Schlegel1998}. Specifically, $\HI$ lines with
Doppler temperature $T_\mathrm{D} \le 1000~\mathrm{K}$ (defined as $22 \times \delta_v^{2}$ where $\delta_v$ 
is the Gaussian linewidth
in $\mathrm{km~s^{-1}}$) 
are associated with gas for which the column density is significantly larger than the optically thin case. 
Their work opens the possibility of identifying optically thin $\HI$ and ``dark gas" 
using $\HI$ line profile information and hence of decomposing the gas phases along the line of sight.
We note that \citet{Kalberla2020} used $\HI$ emission data only to decompose narrow-line $\HI$ gas
and attributed it to the dark gas, hence their results should be validated by an independent way.
We also note that, although 
they
interpreted the dark gas to be primarily
CO-dark $\Htwo$, the thick $\HI$ hypothesis was not ruled out.
To validate their work and establish a method applicable to the Galactic plane, we employed an
$\HI$-line-profile-based analysis to the MBM~53, 54, and 55 clouds and the Pegasus loop. 
They are nearby (100--150~pc) high-latitude clouds \citep{Welty1989, Yamamoto2006} and hence are suitable for the detailed study of the
ISM gas and CRs in the local environment.
This region has been previously studied by \citet{Mizuno2016} using
$\HI$, dust, and $\gamma$-ray data but with a different method.
Specifically, since \citet{Mizuno2016} modeled $\gamma$-ray data using dust maps as a tracer of the total gas column density,
they cannot distinguish optically thick $\HI$ and CO-dark $\Htwo$. Also, the gas-to-dust ratio was calibrated using a small area with high dust temperature,
and hence has large uncertainty. We aim to overcome these difficulties by using $\HI$ linewidth information in this study.

This paper is organized as follows. We describe the properties of the ISM gas templates in Section~2,
and the $\gamma$-ray observations, data selection, and modeling in Section~3. The results of the data analysis are
presented in Section~4, where we confirm that narrow $\HI$ traces dark gas. We also find there remain residuals
and employ a dust map to trace this residual gas. We interpret narrow $\HI$ and residual gas template trace different phases of dark gas,
and discuss the ISM properties in Section~5. We also compare obtained $\gamma$-ray emissivity and CR spectra measured at/near the Earth
and discuss CR properties in Section~5. Finally, a summary of the study and future prospects are presented in Section~6.

\clearpage

\section{ISM Gas Templates and Their Properties}
We analyzed the $\gamma$-ray data in the region of
Galactic longitude $60\arcdeg \le l \le 120\arcdeg$ and
Galactic latitude $-60\arcdeg \le b \le -28\arcdeg$, which encompasses
the MBM 53, 54, and 55 cloud complexes and the Pegasus loop.
We prepared templates of the ISM gas for the $\gamma$-ray data analysis, as we did in \citet{Mizuno2016},
but with updates,
particularly for the atomic-gas phase.
Specifically, we prepared the following gas templates.
All gas maps are stored in a HEALPix \citep{Gorski2005} equal-area sky map of order 9\footnote{
This corresponds to the total number of pixels of $12 \times (2^{9})^{2} = 3145728.$ (9 comes from the resolution index.)
}
with a mean distance of adjacent pixels of $6\farcm9$ ($0.114~\mathrm{deg}$) and pixel size of $0.013~\mathrm{deg^{2}}$. 

\begin{description}
\item[$W_{\HIs}$ maps divided by $\HIalt$ line widths]~\\
We downloaded $\HI$ line profile fits by \citet{Kalberla2020}\footnote{
\url{https://www.astro.uni-bonn.de/hisurvey/AllSky_gauss/}
}
for our region of interest (ROI) and for peripheral regions (${\le}5\arcdeg$ from the boundaries). 
They modeled 
$\HI$ 21-cm emission 
of the HI4PI survey data (with an angular resolution of $16\farcm2$ in full width at the half maximum (FWHM))
in each sky direction using several Gaussians.
And they gave 
the normalization, center, and width of each Gaussian component.
As described in \citet{Kalberla2018}, they required the residuals to be consistent with the noise level,
and also required the number of the used Gaussians as low as possible by considering the information about the 
neighboring pixels.
Negative normalizations or widths are given to suspicious lines (weak lines likely being artifacts due to the noise), 
and we discarded them in preparing the map.
We then divided the $\HI$ data into three components: 
intermediate-velocity clouds (IVCs) \citep[e.g.,][]{Wakker2001}, with velocities outside the range 
from $-30$ to $+20~\mathrm{km~s^{-1}}$ \citep{Fukui2014,Mizuno2016};
$\HI$ clouds with narrow linewidths ($T_\mathrm{D} \le 1000~\mathrm{K}$. 
See \citet{Kalberla2020}. Hereafter, we call them ``narrow $\HI$".);
and those with broad linewidths ($T_\mathrm{D} \ge 1000~\mathrm{K}$; hereafter called ``broad $\HI$").
The $\WHI$ maps (maps of the integrated $\HI$ 21-cm line intensity) of these clouds
are shown in Figure~1.
Using the $\HI$ line profiles in map preparation is a major update over the work by \citet{Mizuno2016}.
In the narrow $\HI$ template, we can recognize coherent structures at around 
$l=84\arcdeg$ to $96\arcdeg$ and $b=-44\arcdeg$ to $-30\arcdeg$
and an area of {$\sim$}$20 \arcdeg \times 20 \arcdeg$ around $(l,b) \sim (109\arcdeg, -45\arcdeg)$.
These features correspond to the MBM~53-55 clouds and the Pegasus loop, respectively.
\item[$\WCO$ map]~\\
As we did in \citet{Mizuno2016}, we used a $\WCO$ map 
(map of the integrated $^{12}$CO (J=1--0) 2.6-mm line intensity)
internally available to the LAT team.
It combines 
the work by \citet{Dame2001} and new data at high Galactic latitudes.
Those data were taken by two 1.2~m telescopes (one in the northern hemisphere and the other in the southern hemisphere)
and smoothed to give an angular resolution of $18\arcmin$ (FWHM) and sampled in $0\fdg25$ intervals.
The new CO data include most of the high-latitude CO clouds in the region studied here.
As described in \citet{Dame2011}, the CO spectra are filtered and integrated over velocities
to suppress noise.
This map also is shown in Figure~1.
\item[\textit{Planck} dust-model maps]~\\
Dust is a known tracer of the total gas column density, and it has been used to construct the dark-gas template
for $\gamma$-ray data analysis \citep[e.g.,][]{Fermi2ndQ,Fermi3rdQ,FermiCham}.
As we did in \citet{Mizuno2016}, we used the \textit{Planck} dust maps 
(of the radiance $R$, the opacity $\tau_{353}$ at 353~GHz, and the dust temperature $T_\mathrm{d}$) of the public Data Release 1
(version R1.20)\footnote{\url{https://irsa.ipac.caltech.edu/data/Planck/release_1/all-sky-maps/}}
described by the \citet{Planck2014}.
As reported in \citet{Mizuno2016}, several areas have high $T_\mathrm{d}$, indicating localized heating by stars. 
To reduce their effects on the $\gamma$-ray data analysis, we refilled these areas
(in the $R$, $\tau_{353}$, and $T_\mathrm{d}$ maps) with the average of the peripheral pixels; see Appendix~A for details.
We also newly employed dust maps of the \textit{Planck} public Data Release 2
(version R2.00)\footnote{\url{https://irsa.ipac.caltech.edu/data/Planck/release_2/all-sky-maps/}} for comparison.
We found that they are less affected by
infrared sources, and we had to mask only one source using the same procedure.
Both Data Release 1 and 2 maps have an angular resolution of $5\arcmin$ (FWHM).

The $R$ and $T_\mathrm{d}$ maps from \textit{Planck} Data Release 1 are shown in Figure~2. We can recognize the MBM~53-55 clouds
and the Pegasus loop in the $R$ map. The $\tau_{353}$ map from \textit{Planck} Data Release 1, and
the $R$ and $\tau_{353}$ maps from the \textit{Planck} Data Release 2 exhibit similar but different contrasts, 
and they predict different
total gas column density distributions. We test them against the $\gamma$-ray data in Section~4.1.
\end{description}

\begin{figure}[htb!]
\gridline{
\fig{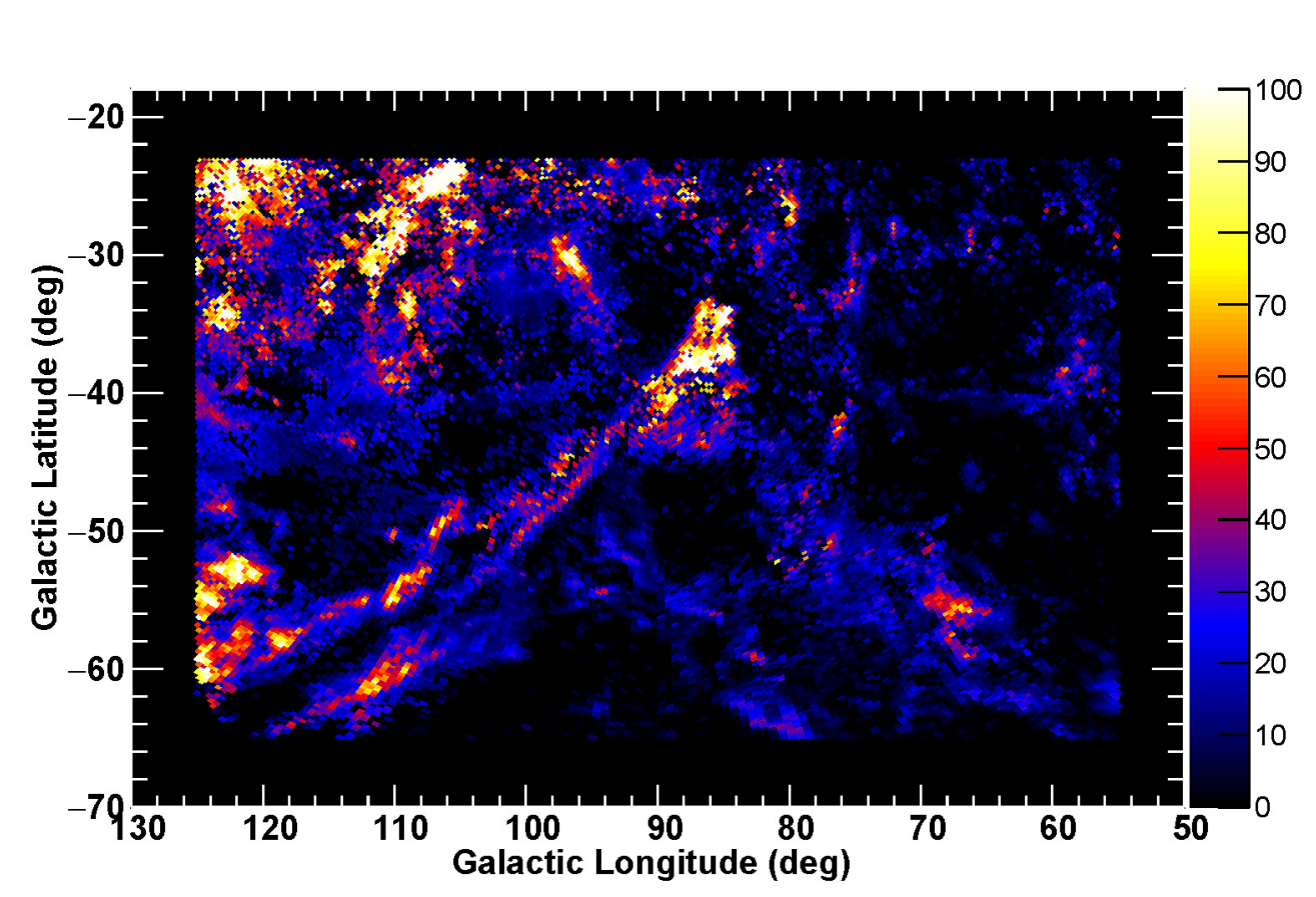}
{0.5\textwidth}{(a)}
\fig{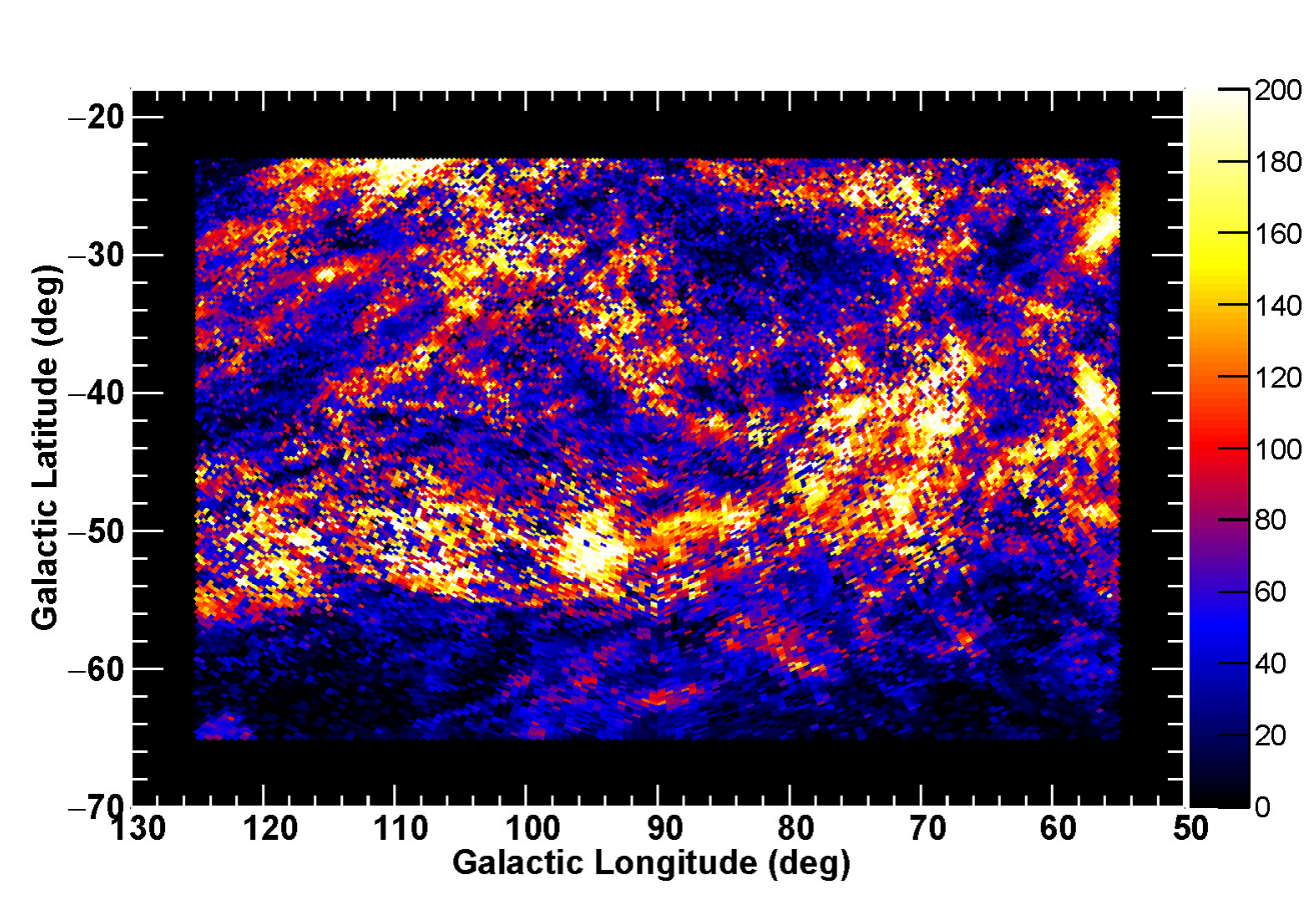}
{0.5\textwidth}{(b)}
}
\gridline{
\fig{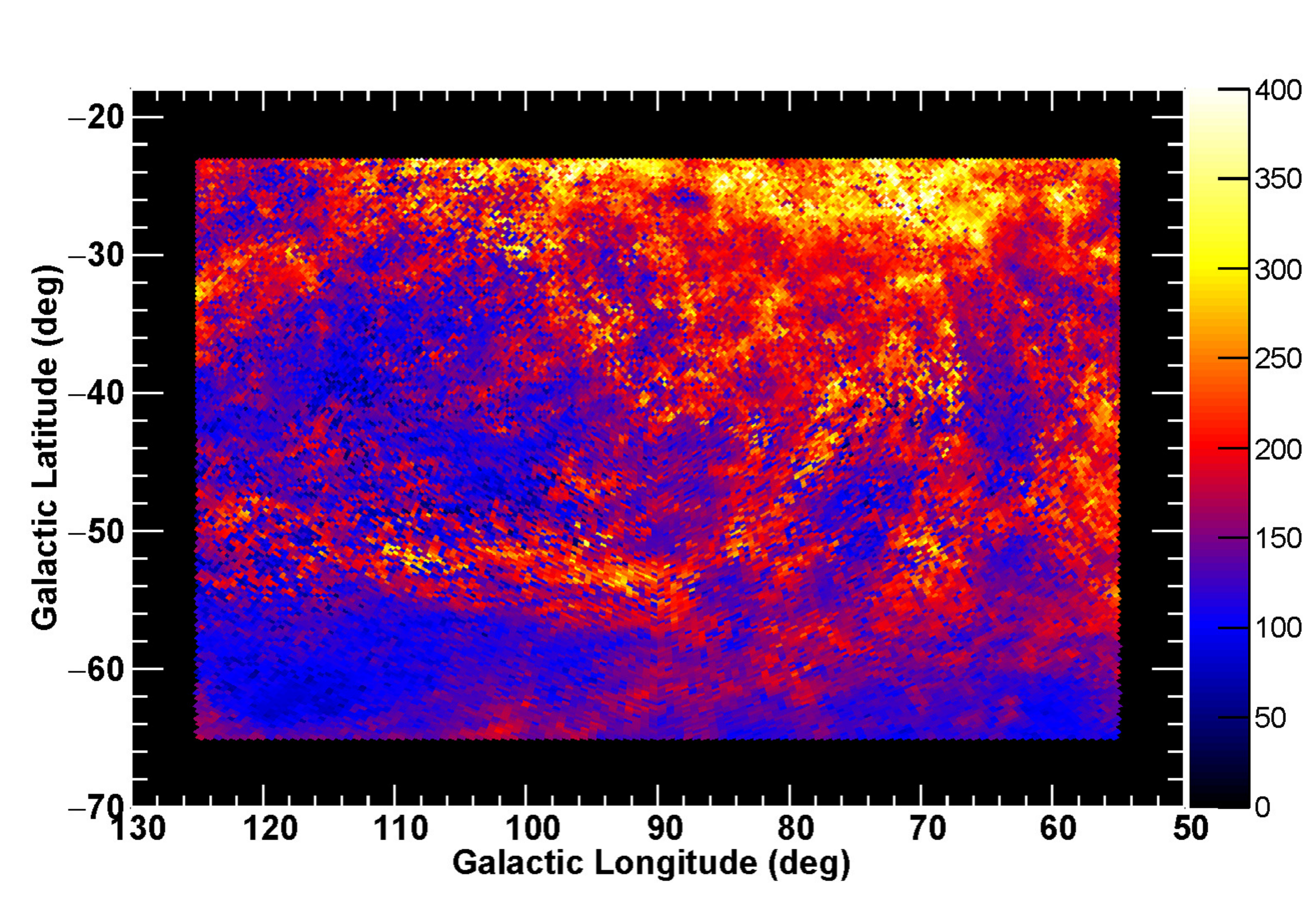}
{0.5\textwidth}{(c)}
\fig{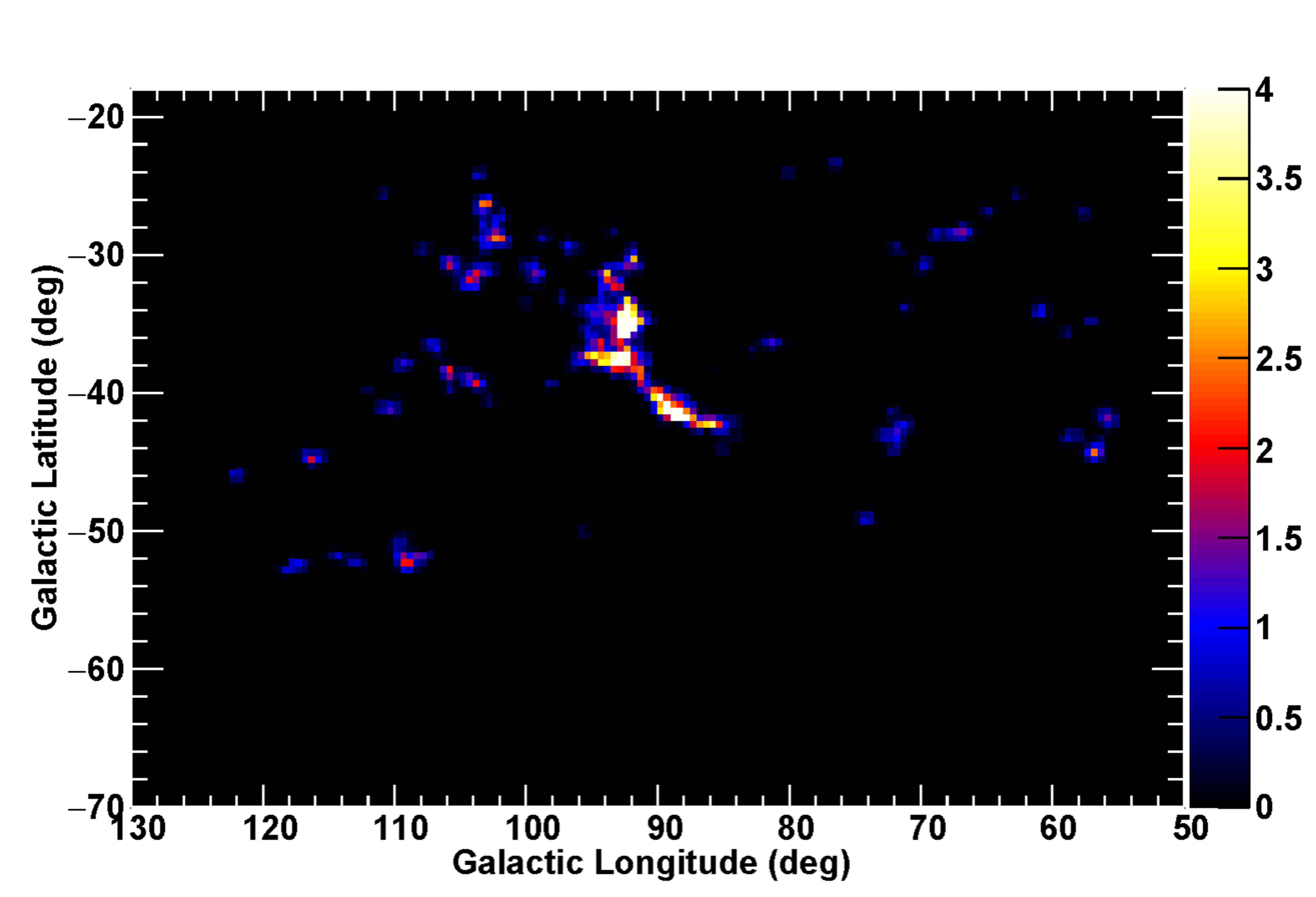}
{0.5\textwidth}{(d)}
}
\caption{
(a) The $\WHI$ map of the IVCs; (b) the $\WHI$ map of narrow $\HI$; (c) the $\WHI$ map of broad $\HI$;
and (d) the $\WCO$ map. All these maps are shown in unit of $\mathrm{K~km~s^{-1}}$.
\label{fig:f1}
}
\end{figure}

\begin{figure}[htb!]
\gridline{
\fig{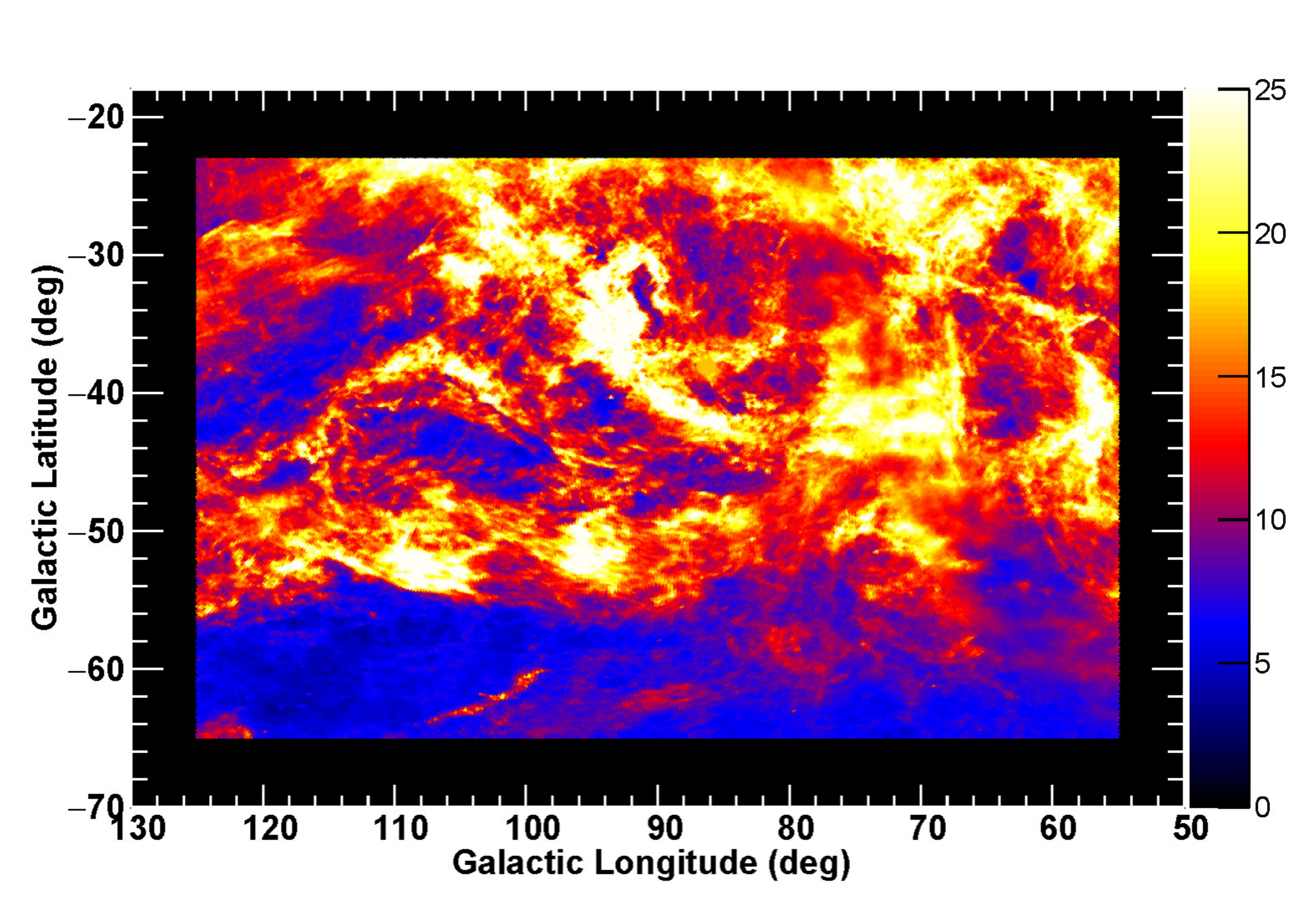}
{0.5\textwidth}{(a)}
\fig{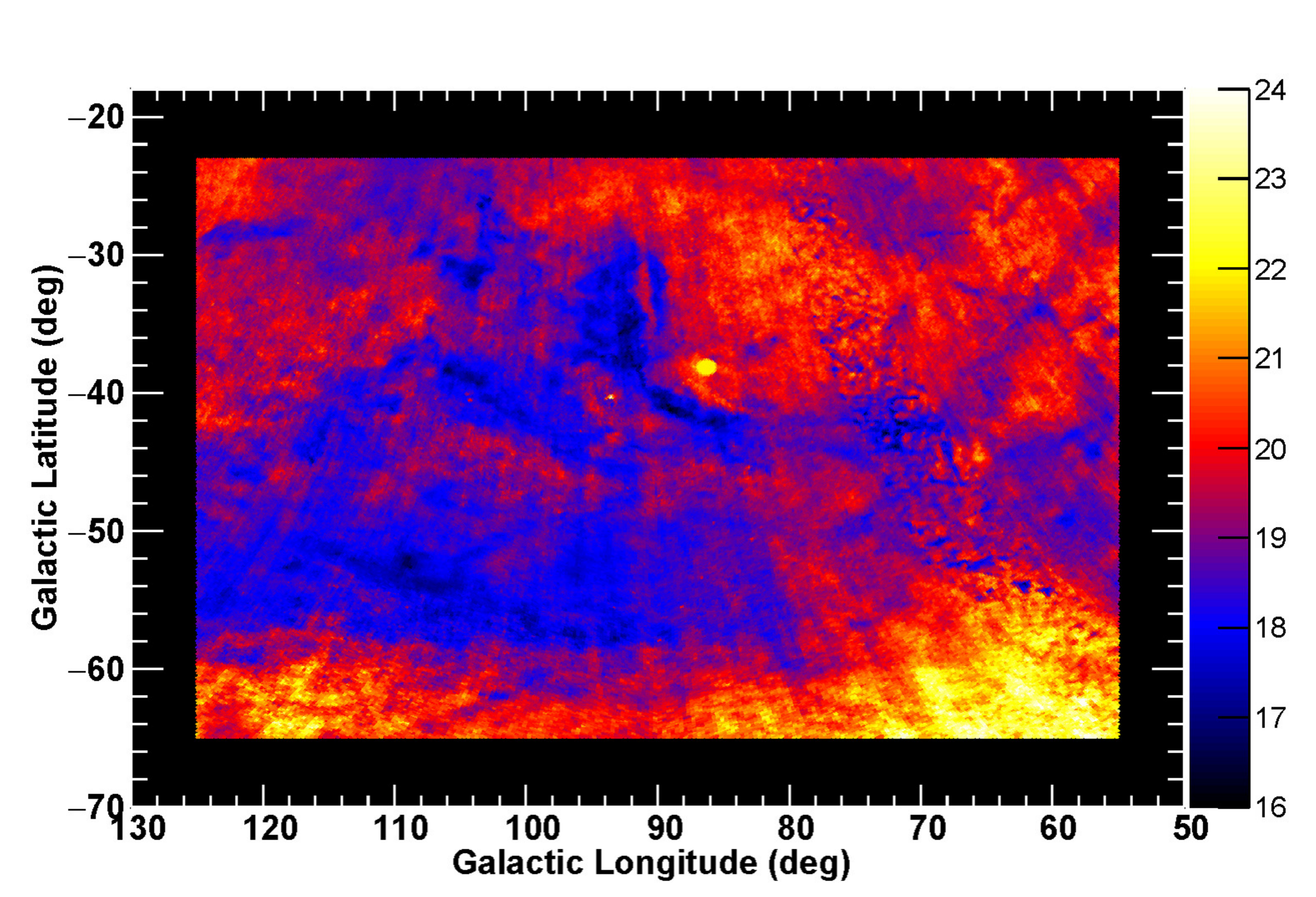}
{0.5\textwidth}{(b)}
}
\caption{
(a) The dust $R$ map ($\mathrm{10^{-8}~W~m^{-2}~sr^{-1}}$); (b) the $T_\mathrm{d}$ map (K). Both maps are from
\textit{Planck} Data Release 1, and the infrared sources are masked as described in Appendix~A.
\label{fig:f2}
}
\end{figure}

\clearpage

\section{Gamma-Ray Data and Modeling}
\subsection{Gamma-ray Observations and Data Selection}
The LAT on board the \textit{Fermi Gamma-ray Space Telescope}, launched in 
2008 June,
is a pair-tracking $\gamma$-ray telescope that detects photons in the range from {$\sim$}20~MeV to more than
300~GeV. Details of the LAT instrument and the pre-launch performance expectations
can be found in \citet{Atwood2009}, and the on-orbit calibration is described in \citet{Abdo2009}.
Thanks to its wide field of view (${\sim}$2.4~sr), \textit{Fermi}-LAT is an ideal telescope to use for studying
Galactic diffuse $\gamma$-rays.
Although the angular resolution is worse than those of gas tracers and energy dependent
(68\% containment radiuses\footnote{
Radius of a circle in which 68\% of photons from a source are contained
} are ${\sim}$5\arcdeg and $0\fdg8$ at 100~MeV and 1~GeV, respectively),
it will be taken into account in the $\gamma$-ray data analysis described in Section~3.3.
Past studies of Galactic diffuse emission by \textit{Fermi}-LAT can be found in, e.g., \citet{FermiPaper2} and \citet{FermiHI2}.

Routine science operations with the LAT started on 
2008 August 4.
We have accumulated events 
from 2008 August 4 to 2020 August 3
(i.e., 12 years) to study diffuse 
$\gamma$-rays in our ROI ($60\arcdeg \le l \le 120\arcdeg$ and $-60\arcdeg \le b \le -28\arcdeg$).
During most of this time interval,
the LAT was operated in sky-survey mode, obtaining complete sky coverage every
two orbits (which corresponds to ${\sim}$3~h), with relatively uniform exposure over time.
We used the latest release of the Pass~8 \citep{Atwood2013,Bruel2018}
data (P8R3), which is less contaminated by a residual background than previous ones.
We used the standard LAT analysis software, \textit{Fermitools}\footnote{
\url{https://fermi.gsfc.nasa.gov/ssc/data/analysis/software/}
}
version 2.0.0,
to select events satisfying the ULTRACLEAN class 
in order to obtain low-background events. 
We also required that the reconstructed zenith angles of the
arrival directions of the photons be less than $100\arcdeg$ and $90\arcdeg$ for energies
above and below 300~MeV, respectively,
to reduce contamination by photons from Earth's atmosphere.
To accommodate the relatively poor angular resolution at low energy, 
we used events and the responses of point-spread-function (PSF) event types 2 and 3 below 300~MeV.
However, above 300~MeV, we did not apply selections based on PSF event types to maximize the photon statistics.
We used the {\tt gtselect} command to apply the selections described above.

In addition, we referred to the Monitored Source List 
light curves\footnote{
\url{https://fermi.gsfc.nasa.gov/ssc/data/access/lat/msl_lc/}
},
and by using the {\tt gtmktime} command,
we excluded the periods ({$\sim$}1600~days in total) during which the LAT detected flares from 3C~454.3.
This significantly reduced contamination in the diffuse emission from the bright active galactic nucleus.
The count-rate threshold was more stringent than that used in \citet{Mizuno2016} in order to
reduce the contamination better.\footnote{
The lists of the mission elapsed time (the number of seconds since 2001 January 1)
that passed the criteria are 2.45--2.72, 3.23--4.17, 5.05--5.45, and larger than 5.47
in $10^{8}$.}
We also excluded the periods during which the LAT
detected bright $\gamma$-ray bursts or solar flares. 
(The integrated time excluded in this procedure is negligible
compared to that excluded to remove data with flares from 3C~454.3.)
Then we prepared a livetime cube by using the {\tt gtltcube} command.
We used the latest response functions that match
our dataset and event selection, P8R3\_ULTRACLEAN\_V3, in the following analysis. 

As described in Section~3.2, we carried out a bin-by-bin likelihood fitting and took account of the energy dispersion.
Using the latest P8R3 data, applying a tighter cut to the 3C454.3 flares, using PSF event types, and taking account of the
energy dispersion are major improvements in data selection and fitting,
allowing us to lower the minimum energy used in the analysis down to 100~MeV.

\subsection{Model to Represent the Gamma-ray Emission}
We modeled the $\gamma$-ray emission observed by \textit{Fermi}-LAT
as a linear combination of the gas column density maps,
IC emission, an isotropic component, and $\gamma$-ray point sources.
The use of the gas column density maps as templates is based on the assumption that 
$\gamma$-rays are generated
via interactions between the CRs and ISM gas and that CR intensities do not vary significantly
over the scale of the interstellar complexes in this study.
This assumption is simple but plausible, particularly in 
high Galactic latitude regions such as the one studied here,
as has been verified by past studies of local clouds using \textit{Fermi}-LAT data
\citep[for a review, see, e.g.,][]{Grenier2015}. 
As described in Section~4, we started with three $\NHI$ maps derived from the $\WHI$ maps
(assuming that the $\HI$ is optically thin) and a $\WCO$ map (see Section~2).
We then improved the templates by using the dust-emission ($D_\mathrm{em}$) model maps.
We used the $\gamma$-ray emissivity model adopted in \citet{FermiHI} to calculate the $\gamma$-ray emissivity.
To model the $\gamma$-rays produced via IC scattering, we used GALPROP\footnote{
\url{http://galprop.stanford.edu}}
\citep[e.g.,][]{Galprop1,Galprop2}.
GALPROP is a numerical code that solves the CR transport equation within the Galaxy and predicts
the $\gamma$-ray emission produced via the interactions of CRs with interstellar matter
and with low-energy photons (IC scattering). 
It calculates the IC emission from the distribution
of propagated electrons and the interstellar radiation field.
Specifically, we utilized the recent work by \citet{Porter2017}.
After testing several IC models against the $\gamma$-ray data with our baseline gas model (three $\NHI$+$\WCO$ templates), 
we decided to use an IC model based on a conventional CR distribution and the ISRF
(SAO-Std model in \citet{Porter2017}); 
see Appendix~B for details.
To model the individual $\gamma$-ray point sources, we referred to the fourth
\textit{Fermi}-LAT catalog (4FGL) described in \citet{Fermi4FGL}, 
which is based on the first 8 years of the science phase of the mission and 
includes more than 5000 sources detected at a significance level of {$\ge$}4$\sigma$. 
For our analysis, we considered 128 4FGL sources (detected at a significance level {$\ge$}5$\sigma$) in our ROI.
In addition, we included bright sources ({$\ge$}20$\sigma$) just outside it (within $10\arcdeg$),
with the parameters fixed at those in the 4FGL,
to consider their possible contamination.
We also added an isotropic component 
to represent the extragalactic diffuse emission
and the residual instrumental background from misclassified CR interactions
in the LAT detector. We adopted the isotropic template provided by the \textit{Fermi} Science Support Center\footnote{
\url{https://fermi.gsfc.nasa.gov/ssc/data/access/lat/BackgroundModels.html}
}.

Then, the $\gamma$-ray intensities $I_{\gamma}(l, b, E)~{\rm(ph~s^{-1}~cm^{-2}~sr^{-1}~MeV^{-1})}$
can be modeled as
\begin{equation}
\begin{split}
I_{\gamma}(l, b, E) & = 
\left[\sum_{i} \CHIi(E) \cdot \NHIi(l, b)  + 
C_\mathrm{CO}(E) \cdot 2X_\mathrm{CO}^{0} \cdot \WCO(l, b) +
C_\mathrm{dust}(E) \cdot X_\mathrm{dust}^{0} \cdot D_\mathrm{em}(l, b) \right] \cdot q_{\gamma}(E) \\
& + C_\mathrm{IC}(E) \cdot I_\mathrm{IC}(l, b, E) +
C_\mathrm{iso}(E) \cdot I_\mathrm{iso}(E) + \sum_{j} {\rm PS}_{j}(l, b, E)~~,
\end{split}
\end{equation}
where the $\NHIi$ is the atomic gas column density (${\rm cm^{-2}}$) model maps,
$q_{\gamma}$ (${\rm ph~s^{-1}~sr^{-1}~MeV^{-1}}$) is the model of the 
differential $\gamma$-ray yield 
or $\gamma$-ray emissivity per H atom,
$\WCO$ is the integrated $^{12}$CO (J=1--0) intensity map ($\mathrm{K~km~s^{-1}}$), $D_\mathrm{em}$ is the dust-emission model
($R$ or $\tau_{353}$), which is a tracer of the total gas column density 
(see Section~4.1) or the residual gas (see Section~4.2).
$R$ and $\tau_{353}$ are given in $\mathrm{W~m^{-2}~sr^{-1}}$ and optical depth, respectively.
The quantities $I_{\rm IC}$ and $I_{\rm iso}$ are the IC model and the isotropic template intensities
(${\rm ph~s^{-1}~cm^{-2}~sr^{-1}~MeV^{-1}}$), respectively, and
${\rm PS}_{j}$ represents the point-source contributions.
The subscript $i$ allows for the use of $\NHI$ maps from separate $\HI$ line profiles.
We adopted the $\gamma$-ray emissivity model used in \citet{FermiHI}.
To accommodate the uncertainties, in either the emissivity model or the gas templates,
we included normalization factors
[$\CHIi$, $C_\mathrm{CO}$, or $C_\mathrm{dust}$ in Equation~(1)] as 
free parameters.
The quantities $X_\mathrm{CO}^{0}$ and $X_\mathrm{dust}^{0}$ are constant scale factors used to make the fitting coefficients
($C_\mathrm{CO}$ and $C_\mathrm{dust}$) close to 1. Specifically, we used
$1 \times 10^{20}~\mathrm{cm^{-2}~(K~km~s^{-1})^{-1}}$ and
$1.82 \times 10^{28}~\mathrm{cm^{-2}~(W~m^{-2}~sr^{-1})^{-1}}$ for
$X_\mathrm{CO}^{0}$ and $X_\mathrm{dust}^{0}$ (for $R$), respectively.
While $\CHIi$ will be 1 if $\NHIi$ represents the true gas column density
and the $\gamma$-ray emissivity agrees with the adopted model,
$C_\mathrm{CO}$ and $C_\mathrm{dust}$ provide the
CO-to-$\Htwo$ conversion factor ($X_\mathrm{CO} \equiv \NHtwo / \WCO$, where $\NHtwo$ gives the
molecular gas column density)
and the dust-to-gas conversion factor, respectively.
$X_\mathrm{CO}$ will be $X_\mathrm{CO}^{0}$ if $\CHIi$ for optically thin $\HI$ 
($\CHIthr$ as will be described in Section~4.1)
and $C_\mathrm{CO}$ are equal
[$X_\mathrm{CO} = X_\mathrm{CO}^{0} \times \left( C_\mathrm{CO}/\CHIthr \right)$].
The IC emission and isotropic models (see above) also are uncertain,
and we have therefore included other
normalization factors [the quantities $C_\mathrm{IC}$ and $C_\mathrm{iso}$ in Equation~(1)] as free parameters.
For the point sources, we adopted spectral models in the 4FGL;
the spectral parameter of normalization was set to be free, and other spectral parameters and 
the positions of each source
were fixed at the values in the 4FGL.
We divided the $\gamma$-ray data into several energy ranges and fitted Equation~(1)
to the $\gamma$-rays in each energy range using the binned-likelihood method, with energy dispersion taken into account,
both implemented in \textit{Fermitools}.

\subsection{Model-Fitting Procedure}
We divided the $\gamma$-ray data into 10 energy bands extending from 0.1 to 72.9~GeV
and stored them in HEALPix maps of order 8 (by using the {\tt gtbin} command).
We used a pixel size a factor of 2 larger than those of the gas maps (Section~2) to accommodate for the small photon statistics,
while keeping the $\gamma$-ray map fine enough to evaluate the gas distribution of $\HI$.
We used energy bins equally spaced logarithmically for the first eight bands 
(e.g., 0.1--0.17, 0.17--0.3, and 0.3-0.52~GeV)
and employed bins twice as broad for the last two bands to accommodate for the small photon statistics.
To evaluate the model spectral shape,
the data were subdivided into three (six) grids within the narrower (broader) energy bands.
Then, in each energy band, 
we prepared exposure and source maps 
with finer grids taken into account (by using the {\tt gtexpcube2} and {\tt gtsrcmaps} commands),
and fitted Equation~(1) to the $\gamma$-ray data
using the binned maximum-likelihood method with Poisson statistics implemented in \textit{Fermitools}
(by importing the {\tt BinnedAnalysis} module in python).
The angular resolution of $\gamma$-ray data is taken into account in this step.
Since the angular resolutions of the gas maps are much
better than that of $\gamma$-ray data in most of the energy range investigated,
gas maps are convolved
with the angular resolution of \textit{Fermi}-LAT
in the $\gamma$-ray data analysis.
Also, since the statistical errors of the $\gamma$-ray data is much larger than the errors of the gas tracer intensity in most pixels, 
we do not take into account the latter in fitting the $\gamma$-ray data.
We modeled $\CHIi$, $C_\mathrm{CO}$, $C_\mathrm{dust}$,
$C_\mathrm{IC}$, and $C_\mathrm{iso}$ as energy-independent normalization factors
in each energy band,
and we modeled $\mathrm{PS}_{j}$ with only the normalization free to vary.

When modeling the point sources, we first sorted them (total 128 sources) by significance
and divided them into 13 groups (each group has 10 sources at the maximum).
We then iteratively fitted them in order of decreasing significance.
First, we fitted the normalizations of the 10 most significant sources;
then, we fitted the normalizations of the second group with parameters of the first group
fixed at the values already determined.
In this way, we worked down to the sources detected at more than 5$\sigma$ in 4FGL.
For each step, the parameters of the diffuse-emission model
were always left free to vary.
After we reached the least-significant sources, we went back
to the brightest ones
and let them and the diffuse-emission models be free to vary while the parameters of the other sources were kept fixed
at the values already determined.
We repeated this process until the increments of the log-likelihoods, $\ln{L}$\footnote{
$L$ is conventionally calculated as $\ln{L}=\sum_{i}n_{i} \ln(\theta_{i})-\sum_{i}\theta_{i}$,
where $n_{i}$ and $\theta_{i}$ are the data and the model-predicted counts in each pixel (for each energy grid)
denoted by the subscript, respectively \citep[see, e.g.,][]{Mattox1996}.
}, 
were less than 0.1 over one loop in each energy band.

\clearpage

\section{Data Analysis}
Most of past $\gamma$-ray analyses used $\HI$ data (single map or maps sorted by velocity), $\WCO$ data, and dust data.
The novelty of this work is to use $\HI$ linewidth information in preparing $\NHI$ maps. Specifically, we prepared $\NHI$ maps of
narrow $\HI$ and broad $\HI$ (Section~2) and assumed that the latter traces optically thin $\HI$.
As will be described in Section~4.1, we expect that narrow $\HI$ traces dark gas and confirm our expectation.
We also find that there remains residual gas and employ a dust map to trace them (Section~4.2).
Final modeling is described in Section~4.3.

\subsection{Initial Modeling}
To examine how well the narrow $\HI$ traces the dark gas,
either optically thick $\HI$ (directly) or CO-dark $\Htwo$ (indirectly),
we started our analysis of $\gamma$-ray data [Figure~3(a)] using the three $\HI$ maps and the $\WCO$ map shown in Figure~1
(hereafter called the ``baseline model") as gas templates.
This is equivalent to setting $C_\mathrm{dust}=0$ in Equation~(1),
while other coefficients ($\CHIi$ and $C_\mathrm{CO}$) are free to vary.
Hereafter, we will use $\CHIone$, $\CHItwo$, and $\CHIthr$ to represent the fit coefficients
for IVC, narrow $\HI$, and broad $\HI$, respectively.
We adopted the optically thin approximation to convert $\WHI$ into $\NHI$.
If broad $\HI$ and narrow $\HI$, respectively, trace the optically thin $\HI$ and the dark gas 
(either optically thick $\HI$ or CO-dark $\Htwo$) well, 
we will have a larger emissivity 
for the latter and a flat fit-residual.
Indeed, we observed that narrow $\HI$ gives about 2 times larger emissivity ($\CHItwo \sim 2 \CHIthr$).
However, we found that our baseline model shows coherent residuals 
\footnote{
Specifically, we refer to residuals at around 
$(l,b) \sim (96\arcdeg, -35\arcdeg)$,
$(105\arcdeg, -38\arcdeg)$, $(103\arcdeg, -40.5\arcdeg)$, 
$(116\arcdeg, -51\arcdeg)$,
and $(108\arcdeg, -54\arcdeg)$. They positionally coincide with the MBM~53-55 clouds and the Pegasus loop seen in narrow
$\HI$ map [Figure~1(b)] and dust $R$ map [Figure~2(a)].
}
in both the MBM~53-55 clouds and the Pegasus loop, as shown in Figure~3(b).
There, we overlaid contours of $R=\mathrm{18 \times 10^{-8}~W~m^{-2}~sr^{-1}}$ to indicate ISM structures.
Considering that \textit{Fermi}-LAT has a very large field of view and has continuously scanned the whole sky
for more than a decade and is well calibrated accordingly
(Section~3.1), such residuals cannot be attributed to the instrumental uncertainty of the $\gamma$-ray data.
This indicates that our baseline model does not fully trace all the neutral gas,
particularly in those ISM structures.
We also applied a $T_\mathrm{D}$-dependent correction proposed by \citet{Kalberla2020} [Equation~(7) in their paper]
to the (narrow + broad $\HI$) map to construct the summed column density map and used it to fit the $\gamma$-ray data. However, we
obtained a residual very similar to that in Figure~3(b).
Therefore, some fraction of gas is missed by $\HI$ 21-cm lines, even if we
adopt the $T_\mathrm{D}$-dependent correction by \citet{Kalberla2020}. 
This motivates us to employ a dust map to model this residual gas.
To determine which dust model to use,
we tested single model maps of the total gas column density ($\NHtot$) proportional to $R$ or $\tau_{353}$ 
(from \textit{Planck} Data Release 1 and 2).
This is equivalent to setting $\CHIi=0$ and $C_\mathrm{CO}=0$.
Although there are fewer degrees of freedom,
each dust map gives a larger value of $\ln{L}$ than does our baseline gas model.
Therefore, we concluded that we can use
a dust map to estimate the residual gas not accounted for by $\HI$ lines or $\WCO$. 
We interpret this residual gas being CO-dark $\Htwo$. 
Since both $\tau_{353}$ and $R$ have potential issues as a tracer of the total 
gas column density \citep[see, e.g.,][and references therein]{Mizuno2016},
we referred to $\ln{L}$ to determine which dust model to use.
We found that the $R$ map from \textit{Planck} Data Release~1 gives the best fit among four "single gas model maps", 
and hence we use it
to construct the residual gas template.
In the following section, we will construct a residual gas template and also examine
if a model with the residual gas template gives a better fit than the model using a single dust map.

\begin{figure}[htb!]
\gridline{
\fig{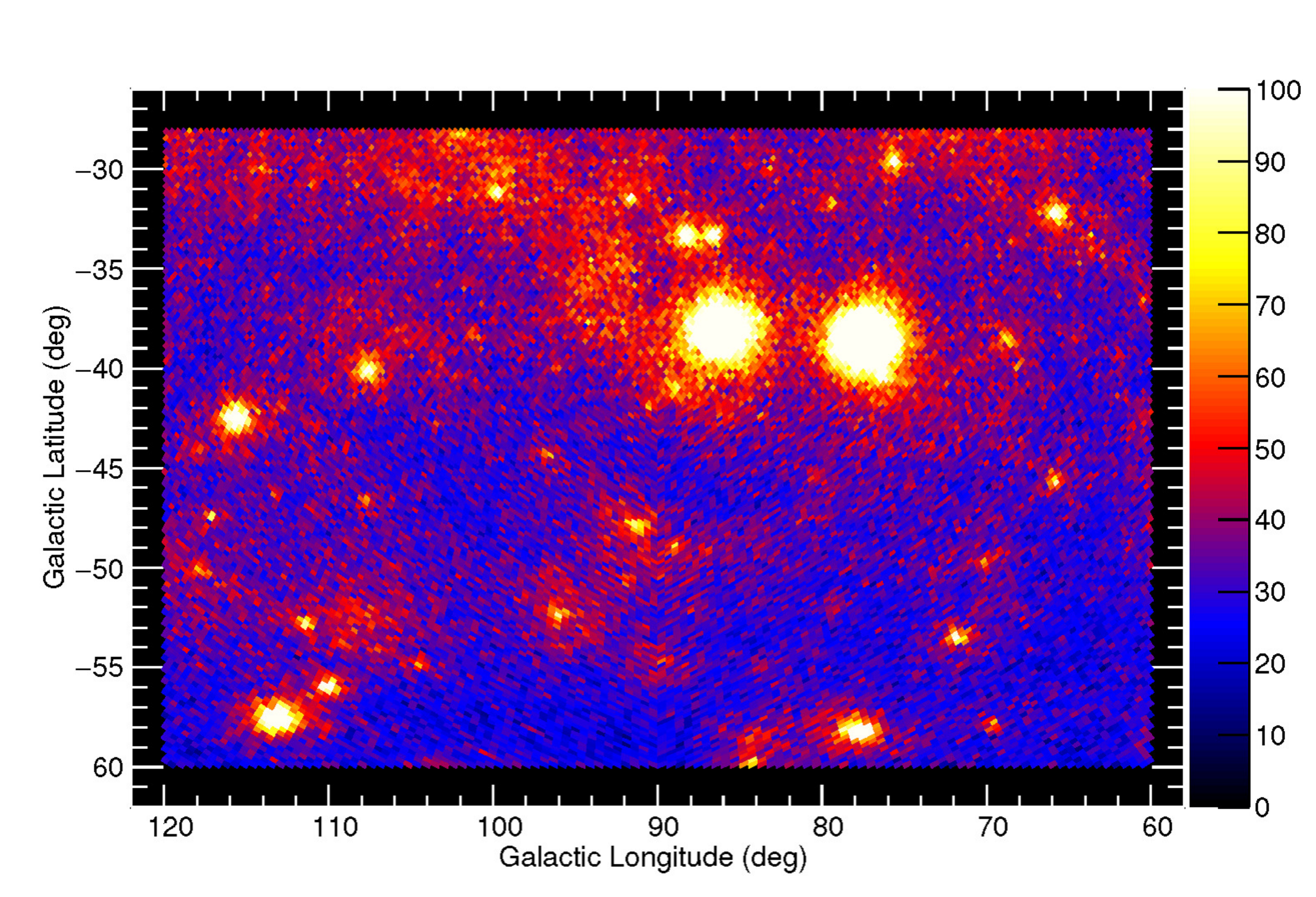}
{0.5\textwidth}{(a)}
\fig{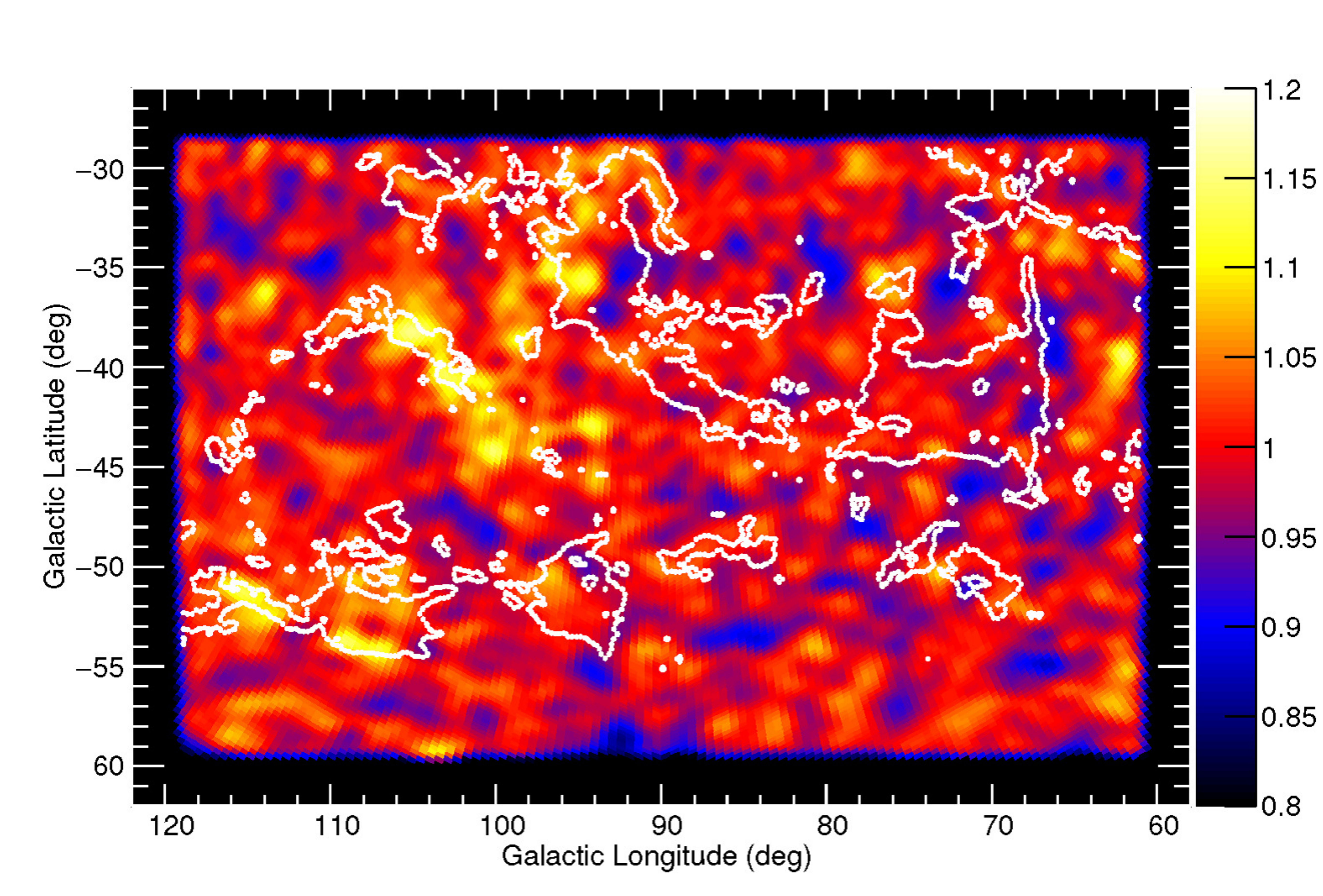}
{0.5\textwidth}{(b)}
}
\caption{
(a) The data count map.
(b) The data/model ratio map using the baseline gas model.
To accommodate small photon statistics (large statistical errors), the panel (b) was smoothed using a 
Gaussian kernel with a standard deviation $\sigma = 60^{'}$.
Contours of $R=\mathrm{18 \times 10^{-8}~W~m^{-2}~sr^{-1}}$ are overlaid to indicate ISM structures.
\label{fig:f3}
}
\end{figure}

\subsection{Fit with the Residual-Gas Template}
Having confirmed that our baseline gas model (the three $\WHI$ maps, divided by using the $\HI$ line profiles, and the $\WCO$ map) is not
good enough to reproduce the $\gamma$-ray data, we add a residual gas template obtained by using
the $R$ map from \textit{Planck} Data Release 1. To construct a good template, we examined the correlations 
among the $\HI$, $\WCO$, and the dust maps in our ROI stored in a HEALPix map of order~9 (see Section~2).
To match the resolution of the
$\WHI$ map, we smoothed the dust maps using a Gaussian kernel with an FWHM of $15\farcm4$.
Since $\WCO$ data has a worse resolution and covers a small fraction of the ROI, we kept the original resolution for it.
Then, by studying the gas properties in detail, we 
removed gas phases other than the residual gas from $\HI$ and $R$ data, and
constructed the residual gas template as follows.
To reduce contamination from CO-bright $\Htwo$, we required $\WCO \le 0.1~\mathrm{K~km~s^{-1}}$,
except in the third step (subtraction of CO-bright $\Htwo$).
Unlike $\HI$ data, we cannot distinguish different gas phases in a dust map along the line of sight using velocity information.
Instead, we use $T_\mathrm{d}$ to best separate gas phases as described below.
We aim to identify optically thin $\HI$ and since $\gamma$-ray data have been generally reproduced
by models using $\HI$ column densities assuming high $T_\mathrm{s}$ ($\ge 125~\mathrm{K}$), $\WCO$, and a residual gas template 
\citep[see, e.g.,][and references therein]{Remy2017},
we assume linearity between $\WHI$ and $R$ at this stage.

\begin{description}
\item[Subtraction of IVCs]~\\
We examined the $\WHI$(total) vs. $R$(total) correlation and found that the outliers in the correlation 
at around $R\ (\mathrm{10^{-8}~W~m^{-2}~sr^{-1}}) \sim 10$ and $\WHI\ (\mathrm{K~km~s^{-1}}) \sim 350$
are affected by IVCs, 
as shown in Figure~4(a).
Most of the pixels with a high IVC fraction (i.e., the fractional $\WHI$ of the IVC is more than {25}\%, represented by the green points) 
exhibit higher $\WHI$/$R$ ratios than the average. 
We understand that this is because IVCs in our ROI are dust poor \citep[e.g.,][]{Fukui2021}. 
While the true $\WHI$/$R$ ratio of the IVCs is uncertain, their contribution to the gas density is small
\citep[integral of $\WHI$ is at the 5\% level; see][]{Mizuno2016}. Therefore, for
simplicity, 
we removed the IVCs from the $\WHI$ and $R$ maps, assuming that they do not have dust (i.e., we subtracted 
the column density of the IVCs from $\WHI$, while keeping the value of $R$ for each pixel).
Unchanging $R$ would overestimate the other gas component, but the effect is minor
since IVCs' contribution is small and their spatial distribution [Figure~1(a)] is
very different from that of other gas phases.
Now we have a $\WHI$ map of narrow and broad $\HI$ (hereafter called $\WHItwothr$).
\item[Subtraction of broad $\HIalt$]~\\
Then we examined the $\WHItwothr$ vs. $R$(total) correlation [Figure~4(b)].
We first aimed to evaluate $R/\WHI$ ratio of broad $\HI$, and
selected areas rich in broad $\HI$ (i.e., with the fractional $\WHI$ of broad $\HI$ more than {95}\%). 
Narrow $\HI$ will have 
a larger $\NH$/$\WHI$ ratio 
(where $\NH$ is the column density of each gas phase) than broad $\HI$ does
if it is optically thick.
CO-dark $\Htwo$ will have an even larger $\NH$/$\WHI$ ratio.
They will have larger $R/\WHI$ ratio, and may not be removed well by the $\WHI$-based selection.
Such an excess gas (over optically thin $\HI$) has been found towards the directions of
low $T_\mathrm{d}$ \citep[e.g.,][]{Mizuno2016, Hayashi2019}.
As described in \citet{Fukui2014}, both the $\HI$ gas and dust are heated by ISRF 
and hence are expected to have a positive correlation between their temperatures. 
Therefore optically thick $\HI$ will be found primarily in the low-$T_\mathrm{d}$ area. CO-dark $\Htwo$ could also have a similar dependence 
on $T_\mathrm{d}.$

Therefore, to further reduce the possible contamination from optically thick $\HI$ and CO-dark $\Htwo$,
we also required these areas to have $T_\mathrm{d} \ge 20~\mathrm{K}$ 
(the red points in the panel). 
We then calculated the average
of $\WHI$ and $R$ in every $40~\mathrm{K~km~s^{-1}}$ bin from $\WHI=100$ to $300~\mathrm{K~km~s^{-1}}$
and obtained the linear relation 
$R\ (\mathrm{10^{-8}~W~m^{-2}~sr^{-1}}) = 0.0463 \cdot \WHI\ (\mathrm{K~km~s^{-1}})$.
We interprete this gives the $R/\WHI$ ratio of broad $\HI$ gas and subtracted them
from the $\WHItwothr$ and $R$(total) maps using this ratio. 
We note that changing the threshold of $T_\mathrm{d}$ by $\pm 1~\mathrm{K}$ affects
the ratio less than 5\%, confirming that a specific choice of the $T_\mathrm{d}$ threshold
does not affect the final map significantly.
Now we have a $\WHI$ map of narrow $\HI$ ($\WHItwo$),
and an $R$ map associated with narrow $\HI$, CO-bright
and the residual gas ($R_\mathrm{2+CO+res}$).
\item[Subtraction of CO-bright $\Htwo$]~\\
We then examined the $\WCO$ and $R_\mathrm{2+CO+res}$ correlation. To select $\WCO$-rich areas, we required $\WCO$ to be greater than $1~\mathrm{K~km~s^{-1}}$.
We then calculated the averages of $\WCO$ and $R$ in every $2~\mathrm{K~km~s^{-1}}$ bin from $\WCO=1$ to 
$12~\mathrm{K~km~s^{-1}}$ (the last bin spans from 9 to $12~\mathrm{K~km~s^{-1}}$)
and fitted them with a linear relation with offset.
Since dark gas lies around CO clouds \citep[e.g.,][]{Grenier2005},
their contribution will be mainly attributed to the offset. We
obtained the linear relation 
$R\ (\mathrm{10^{-8}~W~m^{-2}~sr^{-1}}) = 18.5 + 1.95 \cdot \WCO\ (\mathrm{K~km~s^{-1}})$,
and
subtracted the CO-bright $\Htwo$ gas from the $R_\mathrm{2+CO+res}$ map using the obtained coefficient.
Although there is a significant scatter in Figure~4(c) and hence the uncertainty of $R$/$\WCO$ ratio is large, the effect on the ISM gas properties is small
(except CO-bright $\Htwo$) as will be described in Section~5.1.
Now we have an $R$ map associated with
narrow $\HI$ and the residual gas ($R_\mathrm{2+res}$).
\item[Subtraction of narrow $\HIalt$]~\\
The correlation between the $\WHItwo$ and $R_\mathrm{2+res}$, sorted by $T_\mathrm{d}$ is shown in Figure~4(d).
In order to remove narrow $\HI$ gas, we first aimed to evaluate $R/\WHI$ ratio of it.
In Figure~4(d), we
found that areas with low $T_\mathrm{d}$ exhibit a high $R$/$\WHI$ ratio, which we interpret as being 
the residual gas we found in Section~4.1.
To reduce their contamination, 
we selected high-$T_\mathrm{d}$ areas ($T_\mathrm{d}$ more than 20~K).
We then calculated the average of
$\WHI$ and $R$ in every $30~\mathrm{K~km~s^{-1}}$ bin from $\WHI=0$ to $150~\mathrm{K~km~s^{-1}}$, 
and we obtained the linear relation
$R\ (\mathrm{10^{-8}~W~m^{-2}~sr^{-1}}) = 0.0705 \cdot \WHI\ (\mathrm{K~km~s^{-1}})$.
We removed narrow $\HI$ from the $R_\mathrm{2+res}$ map using this ratio. Now we have the residual gas template 
and we use it in the $\gamma$-ray data analysis.
If $T_\mathrm{s}$ is high throughout $\HI$, as claimed by several past studies, this analysis will give
similar emissivities for broad and narrow $\HI$ templates.
\end{description}

The residual gas template improves the fit significantly; the residuals seen in the MBM 53-55 clouds and the Pegasus loop 
are reduced significantly, and the value of $\ln{L}$ increased by 337.5 with 10 degrees of freedom. 
Therefore the residual gas template
successfully reproduces the gas not traced by our baseline model. 
The model with the residual gas template gives $\Delta \ln{L}=103.4$ with 20 more degrees of freedom
than the model using single $R$ (of \textit{Planck} Data Release~1).
We also found that
narrow $\HI$ gives about 1.5 times larger emissivity than that of broad $\HI$ ($\CHItwo \sim 1.5 \CHIthr$). 
We understand that this is because narrow $\HI$ is
optically thick, and we apply a correction for this in calculating the gas column density.

\begin{figure}[htb!]
\gridline{
\fig{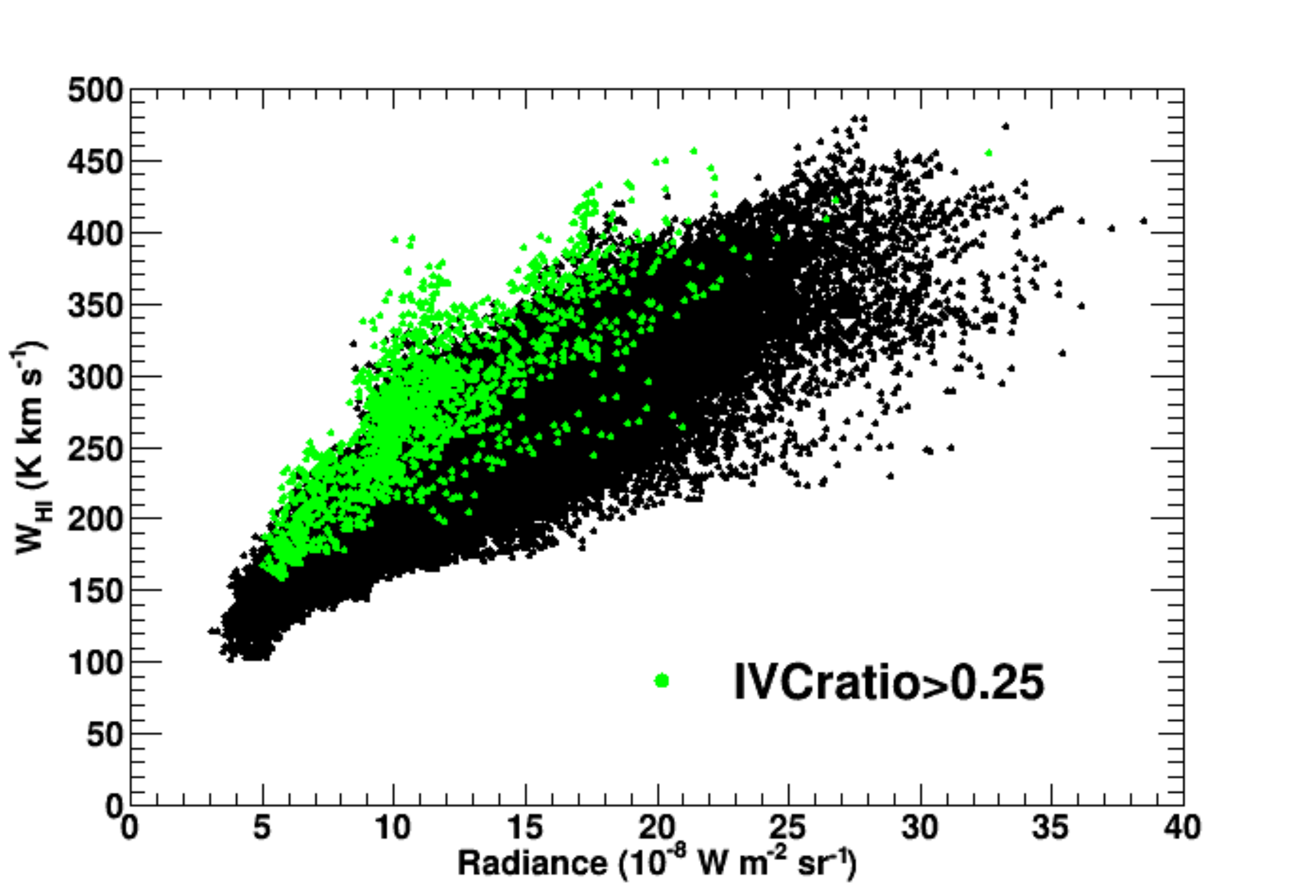}
{0.5\textwidth}{(a)}
\fig{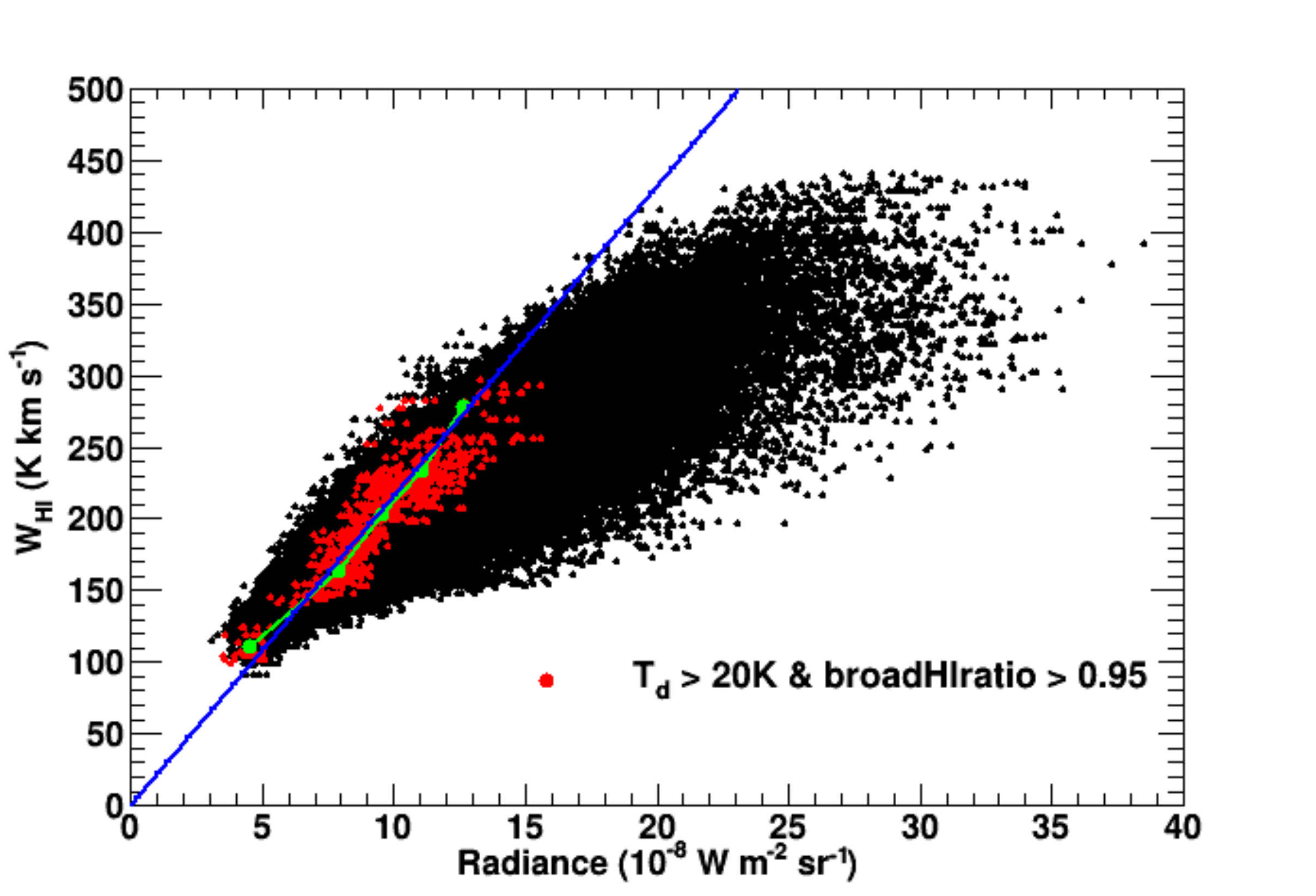}
{0.5\textwidth}{(b)}
}
\gridline{
\fig{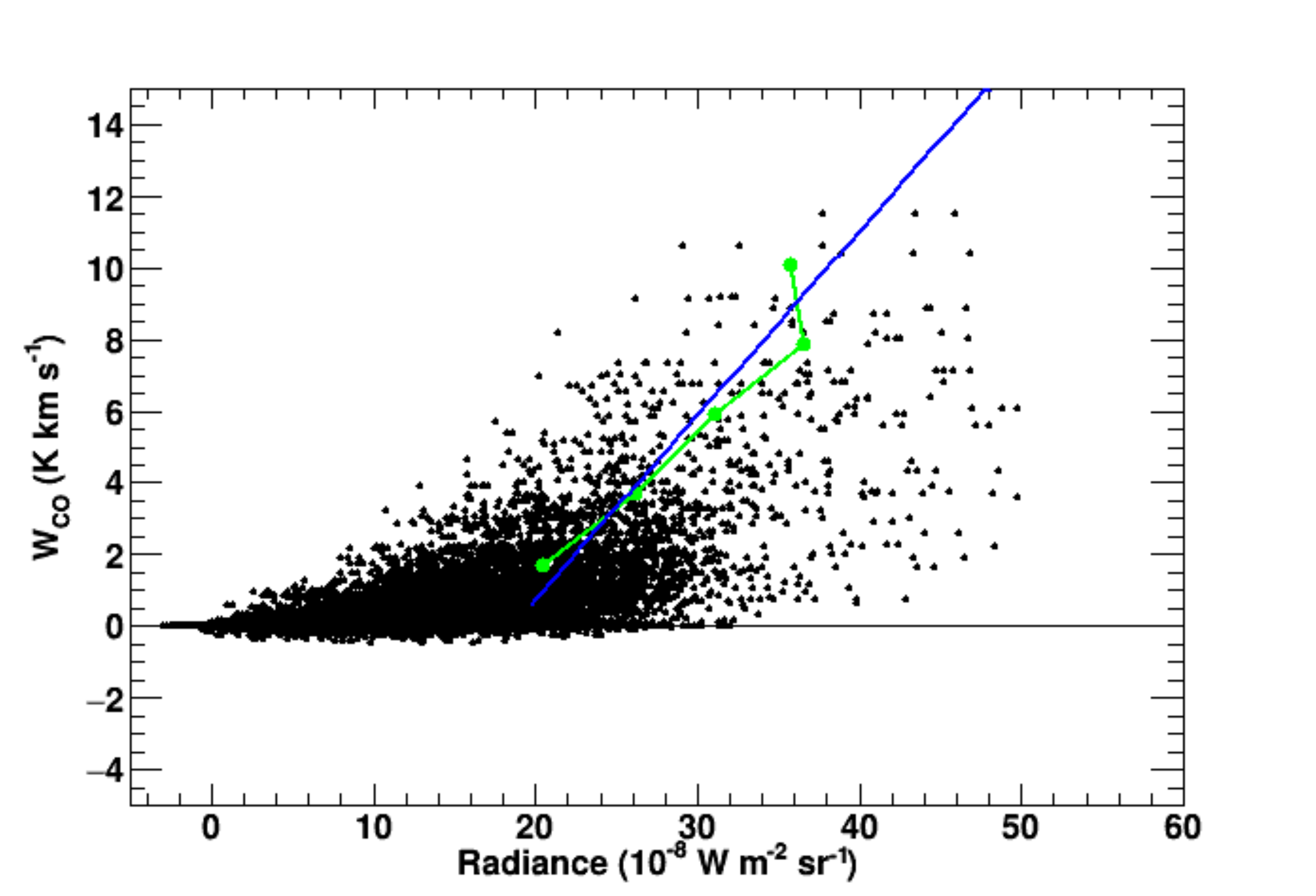}
{0.5\textwidth}{(c)}
\fig{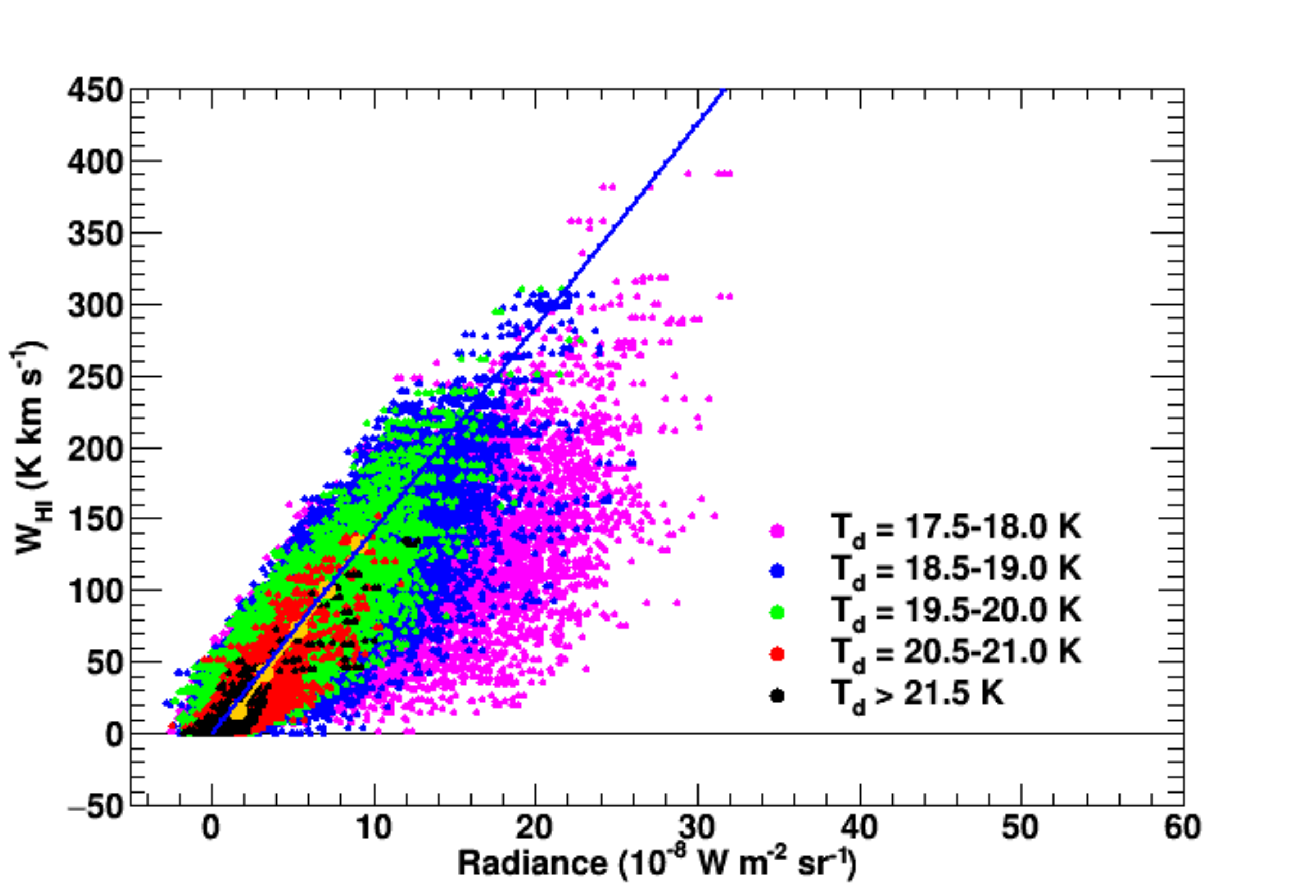}
{0.5\textwidth}{(d)}
}
\caption{
Correlation of ISM gas in our ROI. Each point represents each pixel of our HEALPix map.
(a) The $\WHI$(total)-$R$(total) correlation. Pixels with large IVC fractions are colored green.
(b) The $\WHItwothr$-$R$(total) correlation after subtracting the IVCs. Areas with large
broad $\HI$ fractions and high $T_\mathrm{d}$ are colored red.
The averages of $\WHI$ and $R$ in $\WHI$ bins are shown by green points.
(c) The $\WCO$-$R_\mathrm{2+CO+res}$ correlation after subtracting broad $\HI$.
The averages of $\WCO$ and $R$ in $\WCO$ bins are shown by green points.
(d) The $\WHItwo$-$R_\mathrm{2+res}$ correlation after subtracting CO-bright $\Htwo$.
Data are shown in 0.5~K ranges of
$T_\mathrm{d}$ with 0.5~K gaps between intervals for clarity.
The averages of $\WHI$ and $R$ in $\WHI$ bins are shown by orange points.
In panels (b), (c), and (d), the best fit linear relations are shown by blue lines.
To reduce contamination from CO-bright $\Htwo$, we required $\WCO \le 0.1~\mathrm{K~km~s^{-1}}$
except in panel (c).
See text for details of the procedures used to construct the residual-gas template. 
\label{fig:f4}
}
\end{figure}

\subsection{Final Modeling with $T_\mathrm{s}$ corrections}
We applied a spin-temperature correction to the gas column density
based on \citet{Fukui2014} and \citet{Hayashi2019} to construct a new narrow $\HI$ template and
a residual gas template.
For simplicity, we assumed that the peak brightness temperature ($T_\mathrm{p}$) is
representative of the brightness temperature along the line of sight.
Then, the radiation transfer equation gives
$\WHI$ and the optical depth $\THI$ of the $\HI$ gas as
a function of $T_\mathrm{s}$ and $\Delta \VHI(\equiv \WHI/T_\mathrm{p})$ as
\begin{equation}
\WHI ({\rm K~km~s^{-1}}) = [T_\mathrm{s}(\mathrm{K})-T_\mathrm{bg}(\mathrm{K})]
\cdot \Delta \VHI(\mathrm{km~s^{-1}}) \cdot [1-\exp(-\THI)]~~,
\end{equation}
and 
\begin{equation}
\THI = \frac{\NHI(\mathrm{cm^{-2}})}{1.82 \times 10^{18}}
\cdot \frac{1}{T_\mathrm{s}(\mathrm{K})}
\cdot \frac{1}{\Delta \VHI(\mathrm{km~s^{-1}})}~~,
\end{equation}
where $T_\mathrm{bg}$ is the background continuum radiation temperature 
and $\NHI$ is the gas column density of $\HI$ gas.
Then, we can calculate $\NHI$ as
\begin{equation}
\NHI = -1.82 \times 10^{18} \cdot T_\mathrm{s}(\mathrm{K}) \cdot \Delta \VHI(\mathrm{km~s^{-1}}) \cdot
\log \left[ 1-\frac{\WHI(\mathrm{K~km~s^{-1}})}{[T_\mathrm{s}(\mathrm{K})-T_\mathrm{bg}(\mathrm{K})] \cdot \Delta \VHI(\mathrm{km~s^{-1}})} \right]~~.
\end{equation}
We calculated the value of $\NHI$ for narrow $\HI$ using Equation~(4), assuming a uniform $T_\mathrm{s}$ over the ROI and
$T_\mathrm{bg}=2.7~\mathrm{K}$ (the temperature of the cosmic microwave background radiation).
We note that using a single brightness temperature along the line of sight and a uniform $T_\mathrm{s}$ 
over the ROI is a rough approximation
that may introduce over- or under-prediction at the pixel level.
Also, we compared data and model of Gaussian decomposition for several pixels with a complex profile
\citep[the number of Gaussians more than 7; see][]{Kalberla2018}
and found that parameters of narrow lines could be affected by dominant broad lines there. Therefore,
the value of $T_\mathrm{s}$ should not be taken at face value.
It is a reasonable approach, though, considering the low photon statistics of the $\gamma$-ray data.
Also, it allows us for the first time to apply different $T_\mathrm{s}$ corrections 
to the different $\HI$ gas phases used in the $\gamma$-ray data analysis.

We constructed $\NHI$ maps of narrow $\HI$ assuming $T_\mathrm{s}=120$ to $30~\mathrm{K}$ in 10~K steps.
If $T_\mathrm{p} > T_\mathrm{s}-T_\mathrm{bg}$, Equation~(4) diverges to infinity.
In such a case, we stopped at $T_\mathrm{s}$ that gives $\THI=3$ in Equation~(2),
and calculated $\NHI$ using Equation~(4).
$T_\mathrm{p} > T_\mathrm{s}-T_\mathrm{bg}$ only in a small fraction of pixels (less than 0.1\%) for $T_\mathrm{s} \le 40~\mathrm{K}$.
We then constructed a residual gas template for each $T_\mathrm{s}$.
Specifically, we replaced $\WHI$ in Figure~4(d) with $\NHI(\mathrm{cm^{-2}})$/($1.82 \times 10^{18}$).
Although the difference in $\ln{L}$ is small, we confirmed that
the fit improves gradually as we apply corrections with lower values of $T_\mathrm{s}$ down to $T_\mathrm{s}=40~\mathrm{K}$. 
We therefore concluded that
the templates with a $T_\mathrm{s}$ correction of 40~K applied to narrow $\HI$ represent our best model, and we adopted them as our final model.
The emissivity of narrow $\HI$ now agrees with that of broad $\HI$ at the 10\% level.
Specifically, the averages of $\CHItwo$ (narrow $\HI$) and $\CHIthr$ (broad $\HI$) are $0.980\pm0.018$ and $0.866\pm0.030$, respectively, 
giving the ratio of narrow $\HI$ : broad $\HI$ = 1.13.
The model count map and the data/model ratio map are shown in Figure~5. 
By comparison with Figure~3, we can confirm that
the residuals are reduced significantly.
A summary of the emissivity spectrum and 
the spectrum of each component are presented in Figure~6,
and the best-fit parameters are summarized in Table~1.
Although the spectral shape of CO-bright $\Htwo$ and IVC is apparently different from those of other gas templates,
they are minor components and the errors are large, and hence the effect on other components is small.
Also, although the IC spectrum shows unphysical fluctuation, it anticorrelates with the isotropic spectrum.
Therefore, most of uncertainties of the IC and the isotropic emission (both have smooth spatial distributions)
are mutually absorbed and have small impact on gas emissivities.

\begin{figure}[htb!]
\gridline{
\fig{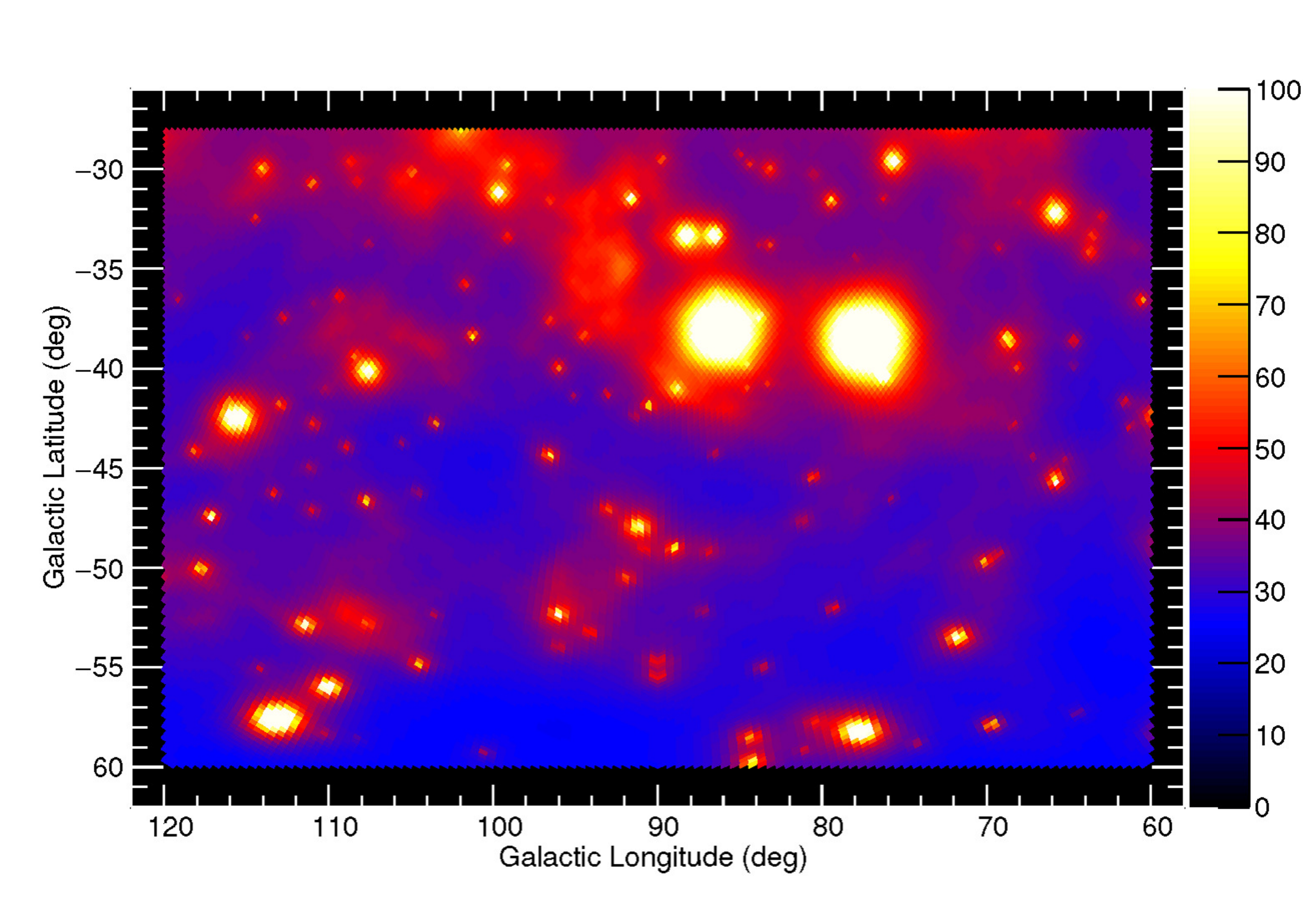}
{0.5\textwidth}{(a)}
\fig{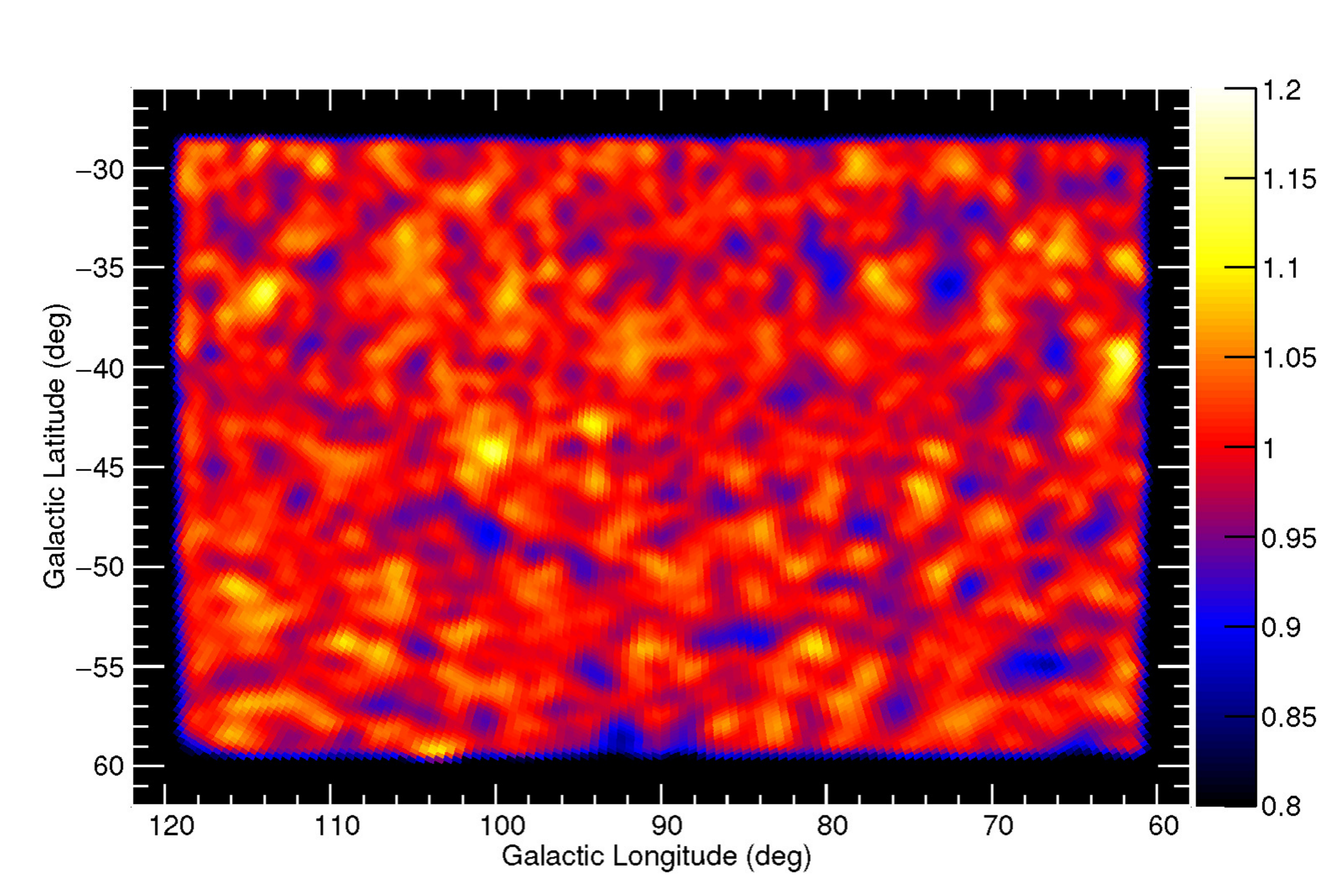}
{0.5\textwidth}{(b)}
}
\caption{
(a) The model count map. (b) The data/model ratio map.
Both maps are obtained with a $T_\mathrm{s}$ correction (of 40~K) applied to the narrow $\HI$ template.
Panel (b) is smoothed using a Gaussian kernel with $\sigma = 60^{'}$ for display.
\label{fig:f5}
}
\end{figure}

\begin{figure}[htb!]
\gridline{
\fig{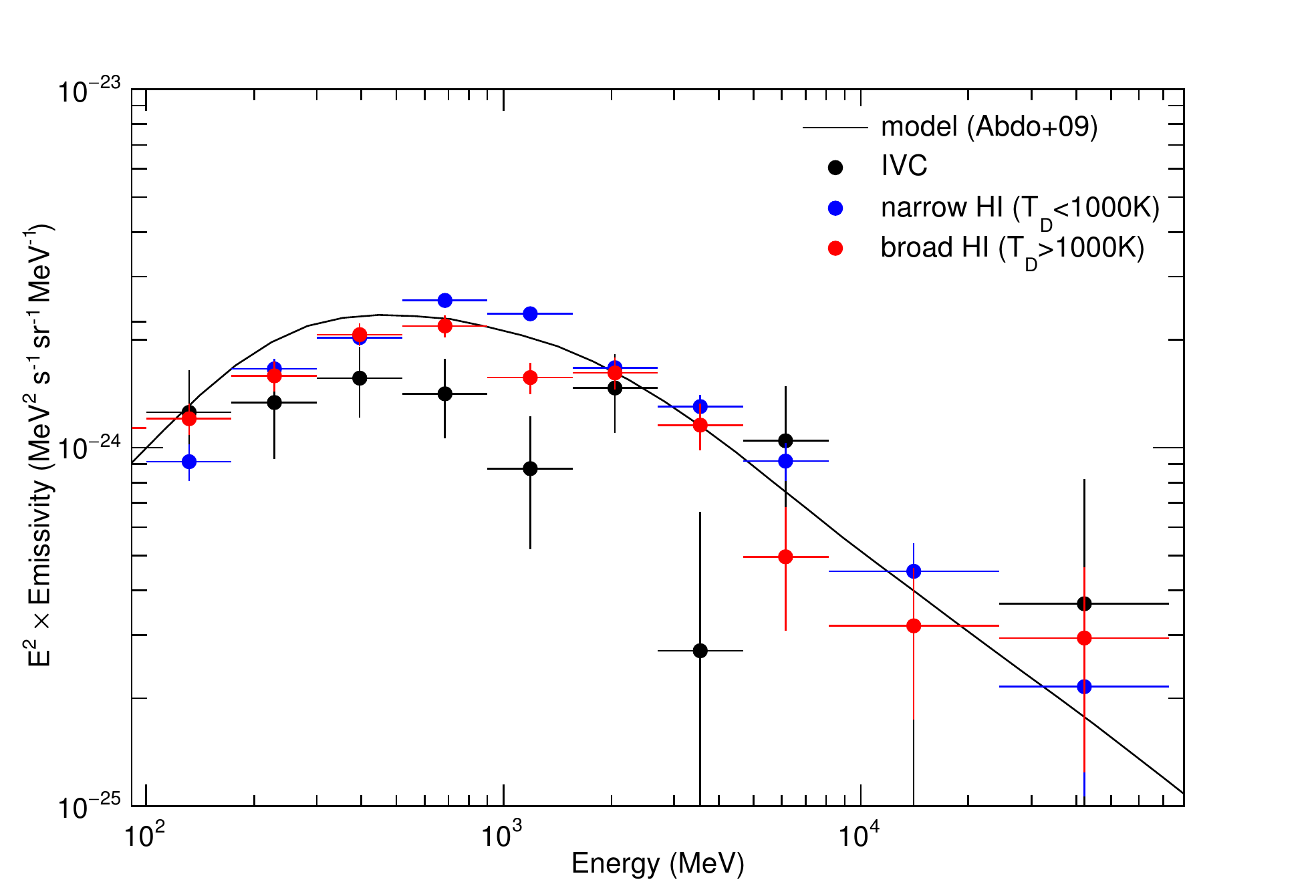}
{0.5\textwidth}{(a)}
\fig{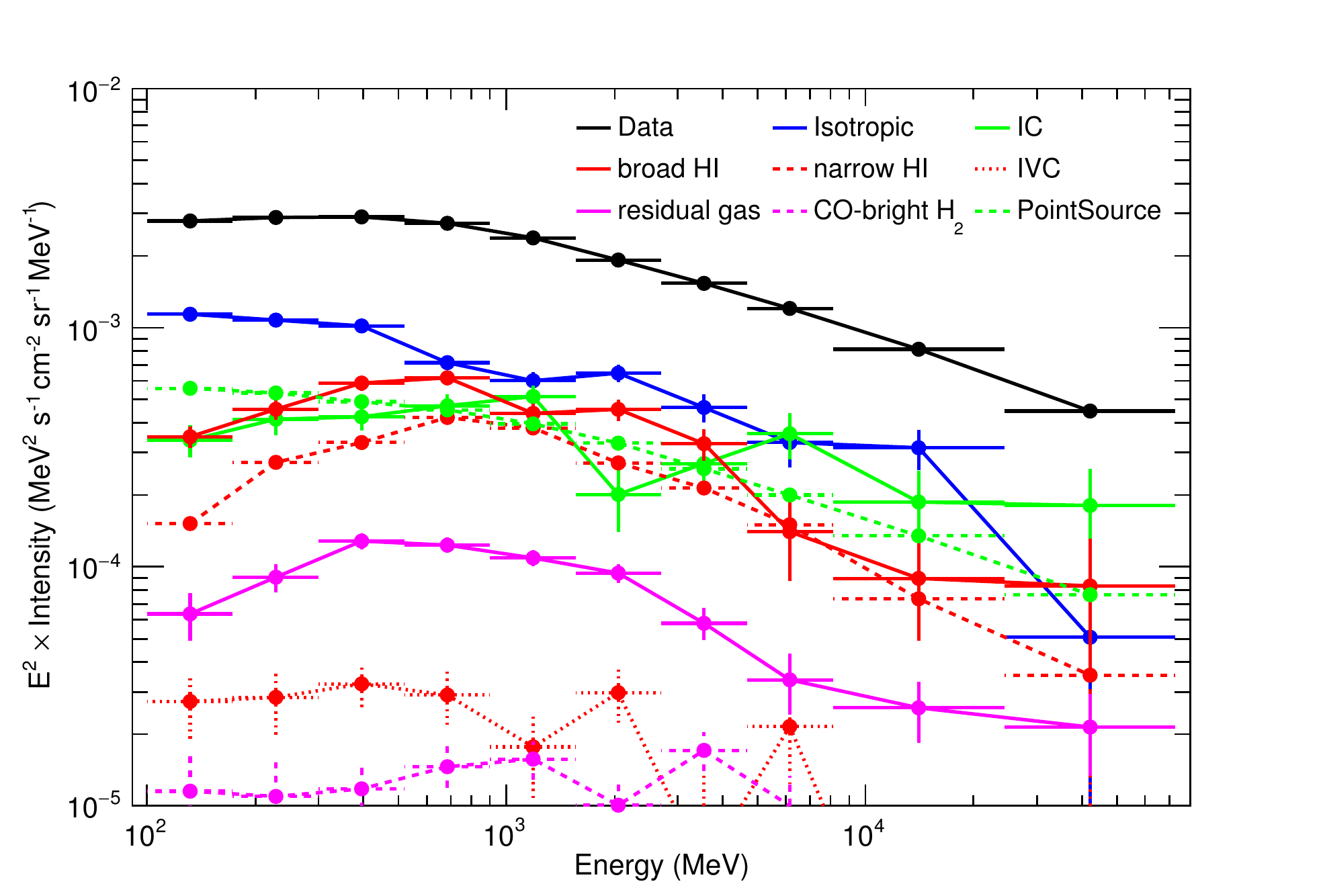}
{0.5\textwidth}{(b)}
}
\caption{
(a) The $\HI$ emissivity spectrum of IVC, narrow $\HI$, and broad $\HI$.
(b) The spectrum of each component. Both panels are obtained by applying $T_\mathrm{s}$ corrections of 40~K to the narrow $\HI$.
\label{fig:f6}
}
\end{figure}

\begin{deluxetable*}{cccccccc}
\tablecaption{Best-fit parameters, with 1-sigma statistical uncertainties,
obtained by using gas templates with a $T_\mathrm{s}$ corrections of 40~K to the narrow $\HI$.}
\tablewidth{0pt}
\tablehead{
\colhead{Energy} & \colhead{$\CHIone$} & \colhead{$\CHItwo$} & \colhead{$\CHIthr$} & \colhead{$C_\mathrm{CO}$} &
\colhead{$C_\mathrm{dust}$} & \colhead{$C_\mathrm{IC}$} & \colhead{$C_\mathrm{iso}$} \\
\colhead{(GeV)} & \colhead{(IVC)} & \colhead{(narrow $\HI$)} & \colhead{(broad $\HI$)} & &
\colhead{} & \colhead{} & \colhead{}
}
\startdata
0.10--0.17 & $0.97 \pm 0.30$ & $0.70 \pm 0.08$ & $0.93 \pm 0.09$ & $0.58 \pm 0.25$ & $0.18 \pm 0.04$ & $0.70 \pm 0.11$ & $1.18 \pm 0.05$ \\
0.17--0.30 & $0.68 \pm 0.21$ & $0.84 \pm 0.05$ & $0.81 \pm 0.08$ & $0.37 \pm 0.15$ & $0.19 \pm 0.03$ & $0.89 \pm 0.13$ & $1.20 \pm 0.06$ \\
0.30--0.52 & $0.66 \pm 0.15$ & $0.86 \pm 0.04$ & $0.88 \pm 0.06$ & $0.34 \pm 0.10$ & $0.22 \pm 0.02$ & $0.99 \pm 0.12$ & $1.25 \pm 0.06$ \\
0.52--0.90 & $0.60 \pm 0.15$ & $1.10 \pm 0.04$ & $0.93 \pm 0.07$ & $0.42 \pm 0.09$ & $0.22 \pm 0.02$ & $1.26 \pm 0.15$ & $1.04 \pm 0.07$ \\
0.90--1.56 & $0.41 \pm 0.17$ & $1.12 \pm 0.04$ & $0.74 \pm 0.08$ & $0.51 \pm 0.09$ & $0.22 \pm 0.02$ & $1.67 \pm 0.18$ & $1.06 \pm 0.09$ \\
1.56--2.70 & $0.86 \pm 0.21$ & $0.98 \pm 0.06$ & $0.95 \pm 0.10$ & $0.40 \pm 0.11$ & $0.24 \pm 0.02$ & $0.79 \pm 0.24$ & $1.44 \pm 0.13$ \\
2.70--4.68 & $0.22 \pm 0.32$ & $1.07 \pm 0.08$ & $0.95 \pm 0.14$ & $0.93 \pm 0.18$ & $0.20 \pm 0.03$ & $1.34 \pm 0.34$ & $1.19 \pm 0.16$ \\
4.68--8.10 & $1.29 \pm 0.54$ & $1.13 \pm 0.14$ & $0.61 \pm 0.23$ & $0.82 \pm 0.27$ & $0.18 \pm 0.05$ & $2.26 \pm 0.49$ & $0.92 \pm 0.20$ \\
8.10--24.3 & $0.13 \pm 0.77$ & $0.97 \pm 0.19$ & $0.69 \pm 0.31$ & $0.46 \pm 0.35$ & $0.24 \pm 0.07$ & $1.66 \pm 0.58$ & $1.06 \pm 0.20$ \\
24.3--72.9 & $1.80 \pm 2.21$ & $1.06 \pm 0.54$ & $1.44 \pm 0.84$ & $0.00 \pm 0.03$ & $0.45 \pm 0.17$ & $2.68 \pm 1.12$ & $0.27 \pm 0.43$ \\
\enddata
\end{deluxetable*}

\clearpage

\section{Discussion}
\subsection{ISM properties}

Assuming uniform CR intensity, we can evaluate the column density of each gas phase.
We tabulated $\NH$ integrated over the ROI in Table~2,
where $\int \NH \,d\Omega$ for broad $\HI$ is calculated assuming the optically thin case.
The corresponding value of narrow $\HI$ is calculated with $T_\mathrm{s}=40~\mathrm{K}$. As described in Section~4.2,
the emissivity ratio of narrow $\HI$ to broad $\HI$ is 1.13.
We interpret this to mean that while the template map of broad $\HI$ represents
optically thin $\HI$ distribution, that of narrow $\HI$ with $T_\mathrm{s}=40~\mathrm{K}$ still underpredicts the
true gas column density. Therefore we took this ratio into account in the calculation 
(i.e, true $\NH$ of narrow $\HI$ is 13\% larger.)
We also calculated the integral of $\NH$ for other gas phases using the average fitting coefficients
given in Table~1. 
Although $R$/$\WCO$ ratio has a large uncertainy (see Section~4.2), the effect on other gas phases is expected to be small since
CO-bright $\Htwo$ is the least significant component in our ROI (Table~2). To examine this expectation, we increased/decreased
the $R$/$\WCO$ ratio by a factor of 2 and repeated the analysis (construction of the residual gas template and 
$\gamma$-ray fit with it). While $C_\mathrm{CO}$ changed by ${\sim}$20\%,
the fit coefficients of other gas components were affected only at the 1\% level.

Because the MBM 53-55 clouds and the Pegasus loop are located at similar distances from the 
solar system, and because most of the $\HI$ clouds are expected to coexist with the $\Htwo$ clouds
(because they are located at high Galactic latitudes),
we can estimate the total mass of gas from $\NH$ as
\begin{equation}
M = \mu m_{\rm H} d^{2} \int \NH \,d\Omega~~,
\end{equation}
where $d$ is the distance to the cloud, $m_{\rm H}$ is the mass of the hydrogen atom,
and $\mu=1.41$ is the mean atomic mass per H atom \citep{AQ2000}.
From Equation~(5),
$\int \NH \,d\Omega=10^{22}~{\rm cm^{-2}~deg^{2}}$ corresponds to {$\sim$}740~${\rm M_{\sun}}$ for
$d=150~{\rm pc}$ \citep{Welty1989}. Therefore the mass of broad $\HI$ is estimated to be ${\sim}3\times10^{4}~\mathrm{M_{\sun}}$.

We interpret broad $\HI$ to be optically thin, narrow $\HI$ to be optically thick, and the residual gas 
to be CO-dark $\Htwo$. The values of $\int \NH \,d\Omega$ for narrow $\HI$ is larger than that of the optically thin case
by $8.0 \times 10^{22}~{\rm cm^{-2}~deg^{2}}$, nearly equal to that of the residual gas. Therefore, 
the ratio of optical depth correction (to the $\HI$ column density)
to CO-dark $\Htwo$ is ${\sim}1$. 
Because of the lack of information on $T_\mathrm{s}$, this value has been
used by several authors as one of possible cases in discussing the ISM properties
\citep[e.g.][]{Planck2015,Remy2018}, and our result supports their assumption.
The fraction of $\HI$ optical depth correction
and CO-dark $\Htwo$,
usually considered dark gas, is about 20\% of the total gas column density. 
This agrees with the value obtained by \citet{Mizuno2016}, which employed a different gas-modeling method.

We summarize the spatial distribution of dark gas in Figure~7.
The left panel shows the distribution of $\HI$ optical depth correction,
and the right panel shows that of CO-dark $\Htwo$ in $\NH$.
They show the different gas distribution and may help us understand 
how the gas evolves from thin $\HI$ to CO-bright $\Htwo$ through
two dark-gas phases (optically thick $\HI$ and CO-dark $\Htwo$).
The CO-dark $\Htwo$ clearly traces the MBM 53-55 clouds and the Pegasus loop well, 
with similar peak $\NH$ in two regions.
On the other hand, $\HI$ optical depth correction (left panel)
shows a similar but less structured distribution, with a larger 
amount of the gas 
in the Pegasus loop.

To examine the gas distributions in more detail, 
we defined two subregions, MBM~53-55 clouds and the Pegasus loop as
$l=84\arcdeg$ to $96\arcdeg$ and $b=-44\arcdeg$ to $-30\arcdeg$ and
$l=99\arcdeg$ to $109\arcdeg$ and $b=-55\arcdeg$ to $-35\arcdeg$, respectively.
Then we calculated the integral of the column density of each gas phase as summarized in Table~3. 
These results give a CO-dark $\Htwo$ fraction, defined as the ratio of CO-dark $\Htwo$ to total $\Htwo$, 
of 0.75 and more than 0.9 for MBM~53-55 clouds and the Pegasus loop, respectively.
These values are larger than typical values (0.3--0.5) obtained by the Planck collaboration 
through dust emission observation in high latitude areas and that by \citet{Wolfire2010} through simulation.
In other words, MBM~53-55 clouds and Pegasus loop were found to be rich in dark gas when normalized  by total $\Htwo$.

A possible explanation of such a large CO-dark $\Htwo$ fraction is due to the small gas column density 
that results in CO photodissociation. The value of $\NH$ we found is
a few $\times 10^{21}~\mathrm{cm^{-2}}$ (Figure~7).
On the other hand,
\citet{Wolfire2010} assumed clouds of larger column density ($\ge 10^{22}~\mathrm{cm^{-2}}$) and
modeled the gas distribution using a photodissociation region code.
We also note that \citet{Smith2014} predict that 
the CO-dark $\Htwo$ fraction anticorrelates to molecular gas column density, 
and $W_\mathrm{CO}$ drops (the CO-dark $\Htwo$ fraction increases) below the column density of a few $\times 10^{21}~\mathrm{cm^{-2}}$. 
Indeed, \citet{Remy2018} reported that the CO-dark $\Htwo$ fraction increases below the column density around this value
in anticenter clouds.

In summary, the amounts of gas in $\HI$ optical depth correction
and CO-dark $\Htwo$ are similar in the Pegasus loop.
The column density of $\Htwo$ is above the threshold (a few $\times 10^{21}~\mathrm{cm^{-2}}$) 
only in thin and filamentary structures, and hence,
molecular gas is predominantly CO-dark. On the other hand, the dark gas in MBM~53-55 clouds is mostly CO-dark
$\Htwo$ phase. The column density is above the threshold in large areas, and hence, 
CO-bright $\Htwo$ starts to form, and CO-dark $\Htwo$ fraction starts to decrease.

We remind that the discussion above depends on the assumption that broad $\HI$ and narrow $\HI$ correspond to
the optically thin $\HI$ and thick $\HI$, respectively. The assumption is based on our findings that
(1) narrow $\HI$ gives {$\sim$}1.5 times larger emissivity than that of broad $\HI$, and
(2) there remains residual gas not accounted for by $\HI$ 21-cm lines. 
The discussion above also depends on our (simplified) optical depth correction
to narrow $\HI$ using Equation~(4).
Systematic and large surveys
of background radio sources will provide measurements of the $\HI$ optical depth and the ultimate answer of gas phases.
Such large surveys, however, may not be feasible to cover all the nearby cloud complexes. Therefore, the analysis using $\gamma$-rays such as the one 
presented here is complementary and worthwhile.
\floattable
\begin{deluxetable}{cc}[htb!]
\tablecaption{
Integral for $\NH$ of each gas phase}
\tablewidth{0pt}
\tablehead{
\colhead{phase} & \colhead{$\int \NH \,d\Omega\ (10^{22}~\mathrm{cm^{-2}~deg^{2}})$}
}
\startdata
broad $\HI$ & 39.9 \\
narrow $\HI$ & 26.1 (8.0 over the thin-$\HI$ case) \\
IVC & 2.8 \\
residual gas & 7.9 \\
CO-bright $\Htwo$ & 1.1 \\
\enddata
\end{deluxetable}

\floattable
\begin{deluxetable}{cccc}[htb!]
\tablecaption{
Integral for $\NH$ $(10^{22}~\mathrm{cm^{-2}~deg^{2}})$ for two subresions}
\tablewidth{0pt}
\tablehead{
\colhead{region} & \colhead{$\HI$ optical depth correction} & \colhead{CO-dark $\Htwo$} & \colhead{CO-bright $\Htwo$} 
}
\startdata
MBM 53-55 & 0.7 & 2.3 & 0.8 \\
Pegasus loop & 1.8 & 1.9 & 0.1 \\
\enddata
\end{deluxetable}

\begin{figure}[htb!]
\gridline{
\fig{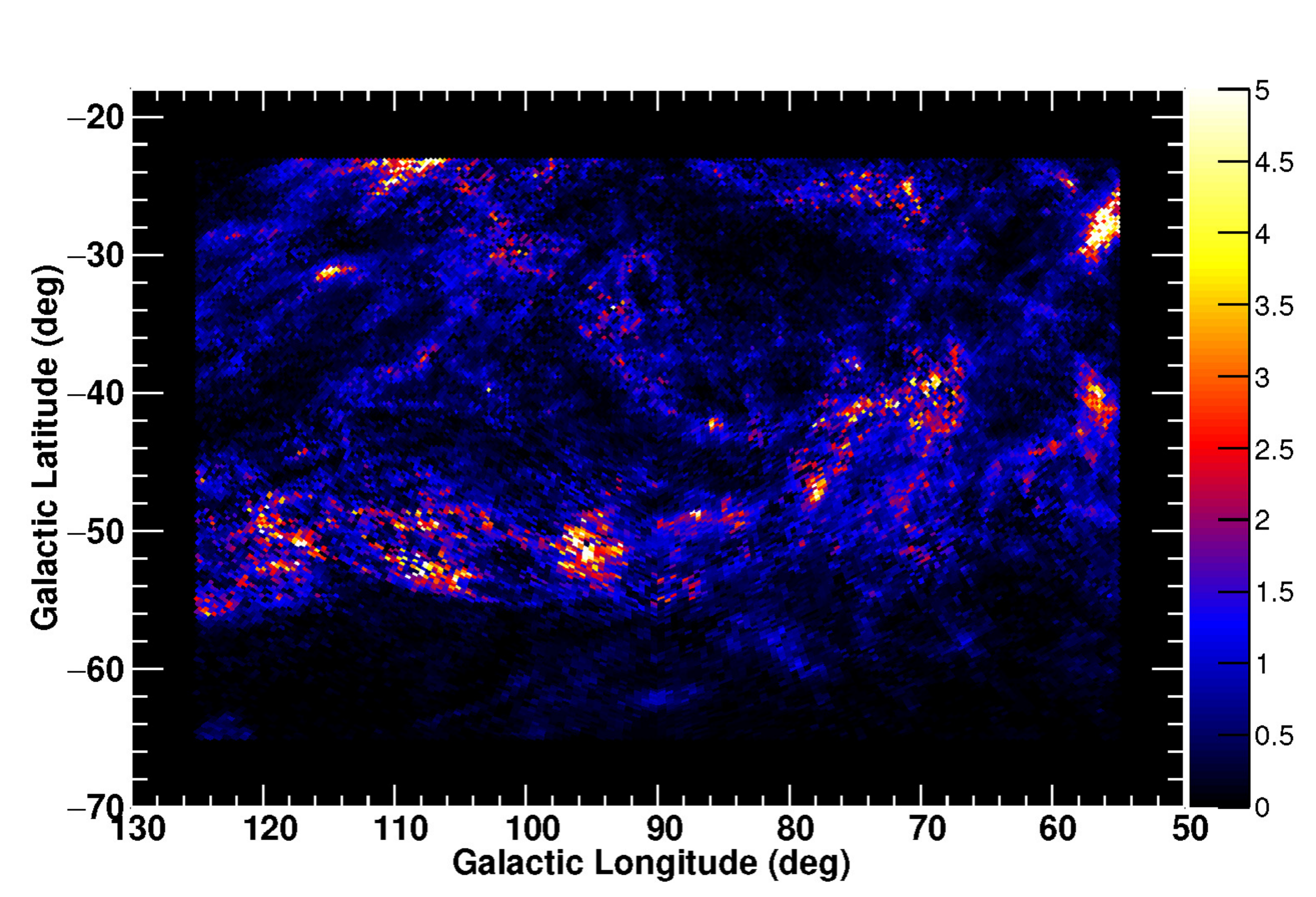}
{0.5\textwidth}{(a)}
\fig{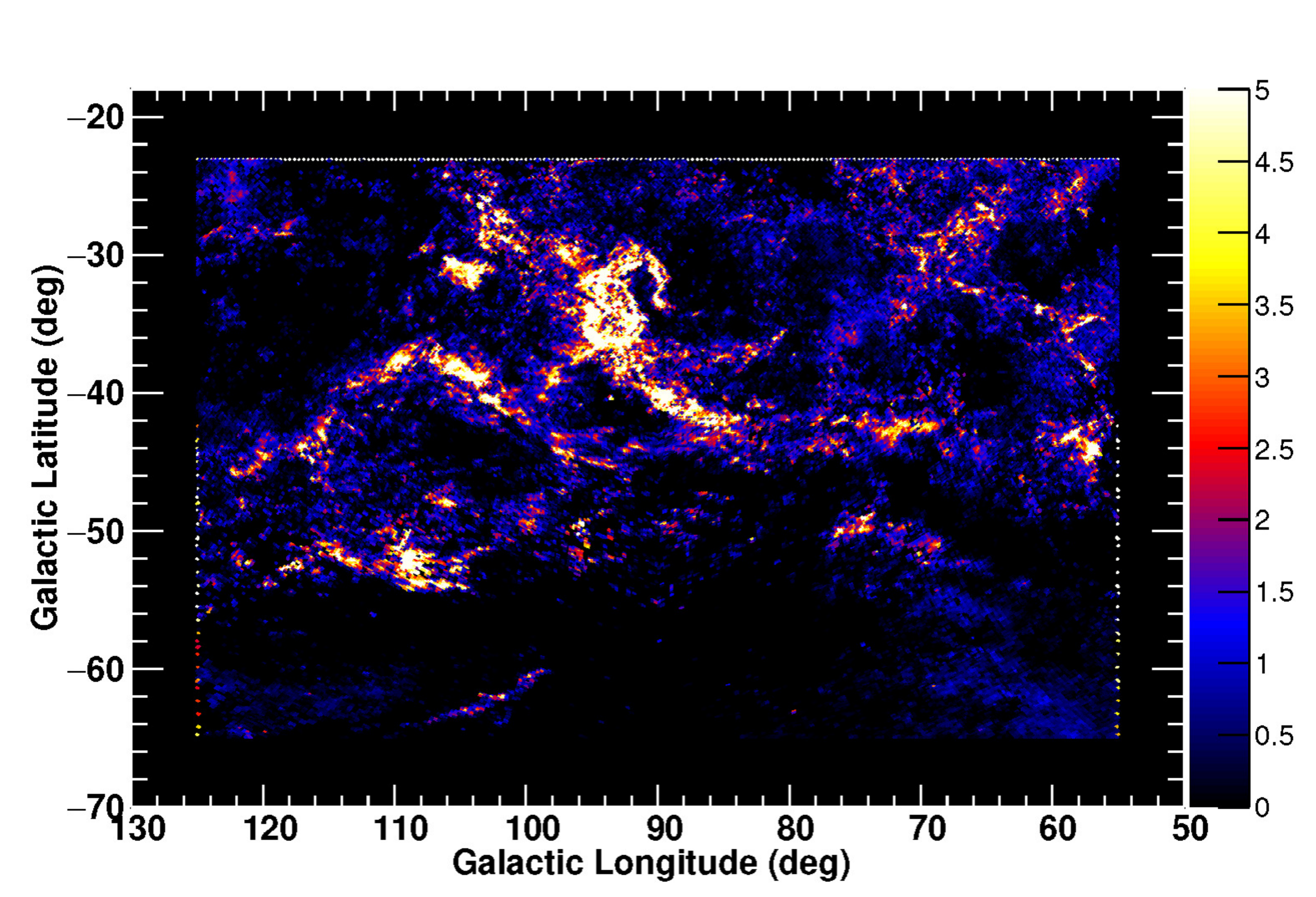}
{0.5\textwidth}{(b)}
}
\caption{
(a) The column density distribution of $\HI$ optical depth correction.
(b) That of CO-dark $\Htwo$. Both distributions are obtained by using the analysis with a $T_\mathrm{s}$ correction of 40~K
to the narrow-$\HI$ template, and shown in $10^{20}~\mathrm{cm^{-2}}$.
\label{fig:f7}
}
\end{figure}

\subsection{CR properties}
Finally, we discuss the $\HI$ emissivity spectrum and the inferred CR spectrum obtained in this study.
Their properties can be evaluated in more detail with fewer gas templates. Therefore,
we added the narrow $\HI$ (with the $T_\mathrm{s}$ correction) and broad $\HI$ templates and re-worked the $\gamma$-ray fitting.
The obtained emissivity spectrum is summarized in Table~4 and Figure~8(a), together with those of other relevant studies
\citep{FermiHI2, Mizuno2020} and the emissivity model 
for the proton local interstellar spectrum (LIS) used in this study.
Since we required the fit coefficient ratio of narrow $\HI$ to broad $\HI$ to be 1.0,
we take the error of the latter ({$\sim$}4\%) as the systematic uncertainty of the emissivity.
We also consider the LAT effective area uncertainty\footnote{
\url{https://fermi.gsfc.nasa.gov/ssc/data/analysis/LAT_caveats.html}};
we assume the uncertainty to be 3\% above 300~MeV, 
and 6\% below 300~MeV where we used only PSF event types 2 and 3.
By adding two types of systematics in quadrature, we obtained the overall uncertainty
to be 5\% and 7\% above and below 300~MeV, respectively.
Even if these systematic uncertainties are taken into account, our emissivity is lower than those of other studies.
Also, while it agrees with the model above 1~GeV, we can recognize a small deviation at low energies.
This suggests that there is a spectral break at
around a few GeV that is stronger than that in the model.
We note that \citet{Mizuno2016} found a hint of a similar deviation, although they could not
give a firm conclusion since their analysis was limited above 300~MeV.

To investigate the inferred CR spectrum in more detail, we carried out simultaneous fitting of
the proton and He CR observations and the $\gamma$-ray emissivity.
We modeled the LIS and the solar-modulation effect using analytical formulae
and then used a Markov chain Monte Carlo (MCMC) technique to constrain the model parameters;
see Appendix~C for details of the framework.
The proton and He LIS models, $J(p)$, are expressed as a power law in momentum ($p$) with two breaks:
\begin{equation}
J(p) \propto \left[ \left( \frac{p}{p_\mathrm{br1}} \right)^{\alpha_{1}/\delta_{1}} 
+ \left( \frac{p}{p_\mathrm{br1}} \right)^{\alpha_{2}/\delta_{1}} \right]^{-\delta_{1}} \cdot 
\left[ 1+\left( \frac{p}{p_\mathrm{br2}} \right)^{\alpha_{3}/\delta_{2}} \right]^{-\delta_{2}}~~.
\end{equation}
There, $\alpha_{1}$ and $\alpha_{2}$ are the indices in the high- and medium-energy ranges, respectively, and 
$p_\mathrm{br1}$ and $\delta_{1}$ control the first (high-energy) spectral break.
It is presumably due to a break in the interstellar diffusion coefficient \citep[e.g.,][]{Ptuskin2006}.
The parameters $p_\mathrm{br2}$ and $\delta_{2}$ control the second break, which represents an
expected break due to ionization \citep[e.g.,][]{Cummings2016},
and $\alpha_{3}$ is the difference in index over this break.
Our formula is motivated by the work by \citet{Strong2015},
but it includes more parameters to represent the CR (and $\gamma$-ray) data
over a broader energy range. 
It is still simple and hence allows us to fit to data within a reasonable computation time.

The $\gamma$-ray emissivity is calculated, based on the proton and He LIS models,
using the AAfrag package \citep{AAfrag} and parameterizations in \citet{Kamae2006}.
We also add an electron/positron bremsstrahlung model by \citet{Orlando2018}, specifically their
best propagation model called PDDE.
In that work, constraints on the electron/positron LIS were obtained 
by fitting the CR direct measurements, the local synchrotron emission from radio to microwaves, and the local $\gamma$-rays emissivity.

In the CR and $\gamma$-ray modeling, we used nine proton datasets, including the AMS-02 data (from 2011 to 2013) and
the Voyager 1 data (from 2012). We also used five He datasets, 
including the AMS-02 and Voyager~1 data from the same periods.
All the CR data is retrieved from the Cosmic-Ray Data Base \citep{Maurin2014}.
We used CR data other than AMS-02 and Voyager~1 to disentangle a possible degeneracy between the
LIS shape and solar modulation. See Appendix~C for details.
We then obtained the parameters of the LIS model as summarized in Table~5.
Other miscellaneous parameters are given in Appendix~C (Table~8).
The inferred CR spectrum and $\gamma$-ray emissivity spectrum are
shown in Figure~8(b) and Figure~8(c), respectively. 
The scaling factor for $\gamma$-rays is $1.07 \pm 0.03$
relative to the AMS-02 spectrum.
Our LIS parameters are primarily constrained by CR data that have very small errors.
The proton spectrum in 1--100~GeV mainly contributes to the $\gamma$-ray emissivity in the
\textit{Fermi}-LAT energy band, and our model emissivity spectrum reproduces the data well
as shown in Figure~8(c).
Our proton spectrum also agrees well with that of \citet{Boschini2020},
which is based on CR data with a detailed calculation of the CR propagation in 
the heliosphere. This supports that our formula [Equation~(6)] and force-field approximation well represent
the LIS and the CR propagation in the heliosphere.
The solar-modulation potential $\Phi$ was found to be about 615~MV for AMS-02.
For the high-energy spectral break, we obtained $p_\mathrm{br1} = 7.1 \pm0.3~\mathrm{GeV}$ and 
$\delta_{1}=0.07\pm0.01$,
confirming earlier claims of this spectral break from $\gamma$-ray data \citep[e.g.,][]{Strong2015}
and CR data \citep[e.g.,][]{Ptuskin2006}.
We note that our break energy is somewhat larger than that inferred from the CR data
using the secondary to primary ratio (usually 3-5~GeV).
This may be due to a spectral break in the CR injection spectrum in addition to
a break in the interstellar diffusion coefficient \citep[e.g.,][]{Galprop2}. 
A detailed comparison of our LIS model to the primary and secondary CR data 
may clarify this issue
and help us better understand the CR acceleration and propagation.

Recently, \citet{Strong2015} and \citet{Orlando2018}
used the high-latitude $\gamma$-ray emissivity spectrum determined by \citet{FermiHI2}
and obtained about 30\% larger proton LIS than that measured at the Earth in the high-energy region. 
The discrepancy is larger than the uncertainties (at the 10\% level) of their studies.
If it is true, the CR spectrum at the Earth is not representative of the LIS. 
To examine this issue we compared proton LIS models in the GeV energy range
based on $\gamma$-ray data (and CR data) in Figure~8(d).
There, our LIS model is multiplied by 1.07 to take account of the scaling factor for $\gamma$-rays,
and shown with the statistical error (3\%) and systematic error (5\%) summed in quadrature
As already shown in Figure~8(a), our emissivity is 10-15\% lower than that by \citet{FermiHI2}, 
giving the LIS model consistent with the AMS-02 spectrum within 10\% in high energy where the solar modulation is irrelevant.
Therefore, while the previous studies required the LIS larger than
the PAMELA/AMS-02 spectra by {$\sim$}30\%, our new study shows better agreement. 
We note that while \citet{FermiHI2} carried out detailed modeling of the high-latitude region and provided a very 
precise measurement of the $\gamma$-ray emissivity spectrum, 
it used Pass~7 data and assumed a uniform
$T_\mathrm{s}$, and hence the obtained spectrum might suffer from
bias on the absolute value.
Also, it samples a region of around {$\sim$}1~kpc and hence may deviate from the prediction based on  the 
CR spectrum directly measured at the Earth, while the MBM~53-55 clouds and the Pegasus loop are very close to us.
Of course, our result is based on a particular region in the sky, and a systematic study of local regions is crucial to settle the issue.
Such a systematic study is also crucial to investigate a possible local variation of the CR spectrum.

\begin{deluxetable*}{cchhhhhh}
\tablecaption{The emissivity multiplised by $E^{2}$}
\tablewidth{0pt}
\tablehead{
\colhead{Energy} & \colhead{$E^{2} \times \mathrm{Emissivity}$} & \nocolhead{$\CHItwo$} & \nocolhead{$\CHIthr$} & \nocolhead{$C_\mathrm{CO}$} &
\nocolhead{$C_\mathrm{dust}$} & \nocolhead{$C_\mathrm{IC}$} & \nocolhead{$C_\mathrm{iso}$} \\
\colhead{(GeV)} & \colhead{($\mathrm{10^{-24}~MeV^{2}~s^{-1}~sr^{-1}~MeV^{-1}}$)} & \nocolhead{} & \nocolhead{} & &
\nocolhead{} & \nocolhead{} & \nocolhead{}
}
\startdata
0.10--0.17 & $1.00 \pm 0.06$ & $0.70 \pm 0.08$ & $0.93 \pm 0.09$ & $0.58 \pm 0.25$ & $0.18 \pm 0.04$ & $0.70 \pm 0.11$ & $1.18 \pm 0.05$ \\
0.17--0.30 & $1.51 \pm 0.07$ & $0.84 \pm 0.05$ & $0.81 \pm 0.08$ & $0.37 \pm 0.15$ & $0.19 \pm 0.03$ & $0.89 \pm 0.13$ & $1.20 \pm 0.06$ \\
0.30--0.52 & $1.86 \pm 0.06$ & $0.86 \pm 0.04$ & $0.88 \pm 0.06$ & $0.34 \pm 0.10$ & $0.22 \pm 0.02$ & $0.99 \pm 0.12$ & $1.25 \pm 0.06$ \\
0.52--0.90 & $2.26 \pm 0.07$ & $1.10 \pm 0.04$ & $0.93 \pm 0.07$ & $0.42 \pm 0.09$ & $0.22 \pm 0.02$ & $1.26 \pm 0.15$ & $1.04 \pm 0.07$ \\
0.90--1.56 & $1.98 \pm 0.07$ & $1.12 \pm 0.04$ & $0.74 \pm 0.08$ & $0.51 \pm 0.09$ & $0.22 \pm 0.02$ & $1.67 \pm 0.18$ & $1.06 \pm 0.09$ \\
1.56--2.70 & $1.51 \pm 0.07$ & $0.98 \pm 0.06$ & $0.95 \pm 0.10$ & $0.40 \pm 0.11$ & $0.24 \pm 0.02$ & $0.79 \pm 0.24$ & $1.44 \pm 0.13$ \\
2.70--4.68 & $1.15 \pm 0.08$ & $1.07 \pm 0.08$ & $0.95 \pm 0.14$ & $0.93 \pm 0.18$ & $0.20 \pm 0.03$ & $1.34 \pm 0.34$ & $1.19 \pm 0.16$ \\
4.68--8.10 & $0.74 \pm 0.09$ & $1.13 \pm 0.14$ & $0.61 \pm 0.23$ & $0.82 \pm 0.27$ & $0.18 \pm 0.05$ & $2.26 \pm 0.49$ & $0.92 \pm 0.20$ \\
8.10--24.3 & $0.38 \pm 0.07$ & $0.97 \pm 0.19$ & $0.69 \pm 0.31$ & $0.46 \pm 0.35$ & $0.24 \pm 0.07$ & $1.66 \pm 0.58$ & $1.06 \pm 0.20$ \\
24.3--72.9 & $0.21 \pm 0.09$ & $1.06 \pm 0.54$ & $1.44 \pm 0.84$ & $0.00 \pm 0.03$ & $0.45 \pm 0.17$ & $2.68 \pm 1.12$ & $0.27 \pm 0.43$ \\
\enddata
\end{deluxetable*}

\floattable
\begin{deluxetable}{ccc}[htb!]
\tablecaption{
Parameters of the LIS obtained by fitting the CR and $\gamma$-ray data}
\tablecolumns{5}
\tablewidth{0pt}
\tablehead{
\colhead{parameter} & \colhead{proton} & \colhead{He}
}
\startdata
normalization & $26.20 \pm 0.10$ & $1.41 \pm 0.01$ \\
$\alpha_{1}$ & $2.86 \pm 0.01$ & $2.76 \pm 0.01$ \\
$\alpha_{2}$ & \multicolumn{2}{c}{$2.57 \pm 0.02$} \\
$\alpha_{3}$ & $-3.25 \pm 0.05$ & $-2.73 \pm 0.08$ \\
$R_{br1}$ (GV) & \multicolumn{2}{c}{$7.08 \pm 0.31$} \\
$\delta_{1}$ & \multicolumn{2}{c}{$0.07 \pm 0.01$} \\
$R_{br2}$ (GV) & $0.524 \pm 0.015$ & $1.314 \pm 0.061$ \\
$\delta_{2}$ & $1.91\pm0.07$ & $0.82 \pm 0.07$ \\
\enddata
\tablecomments{
Normalization is the flux [$\mathrm{c~s^{-1}~m^{-2}~sr^{-1}~(GeV/n)^{-1}}$] at $p = 10~\mathrm{GeV}$.
CR protons and He are assumed to share the value of the high-energy spectral break in rigidity.
Therefore breaks in rigidity ($R_\mathrm{br}$) are given instead of breaks in momentum.
}
\end{deluxetable}

\begin{figure}[htb!]
\gridline{
\fig{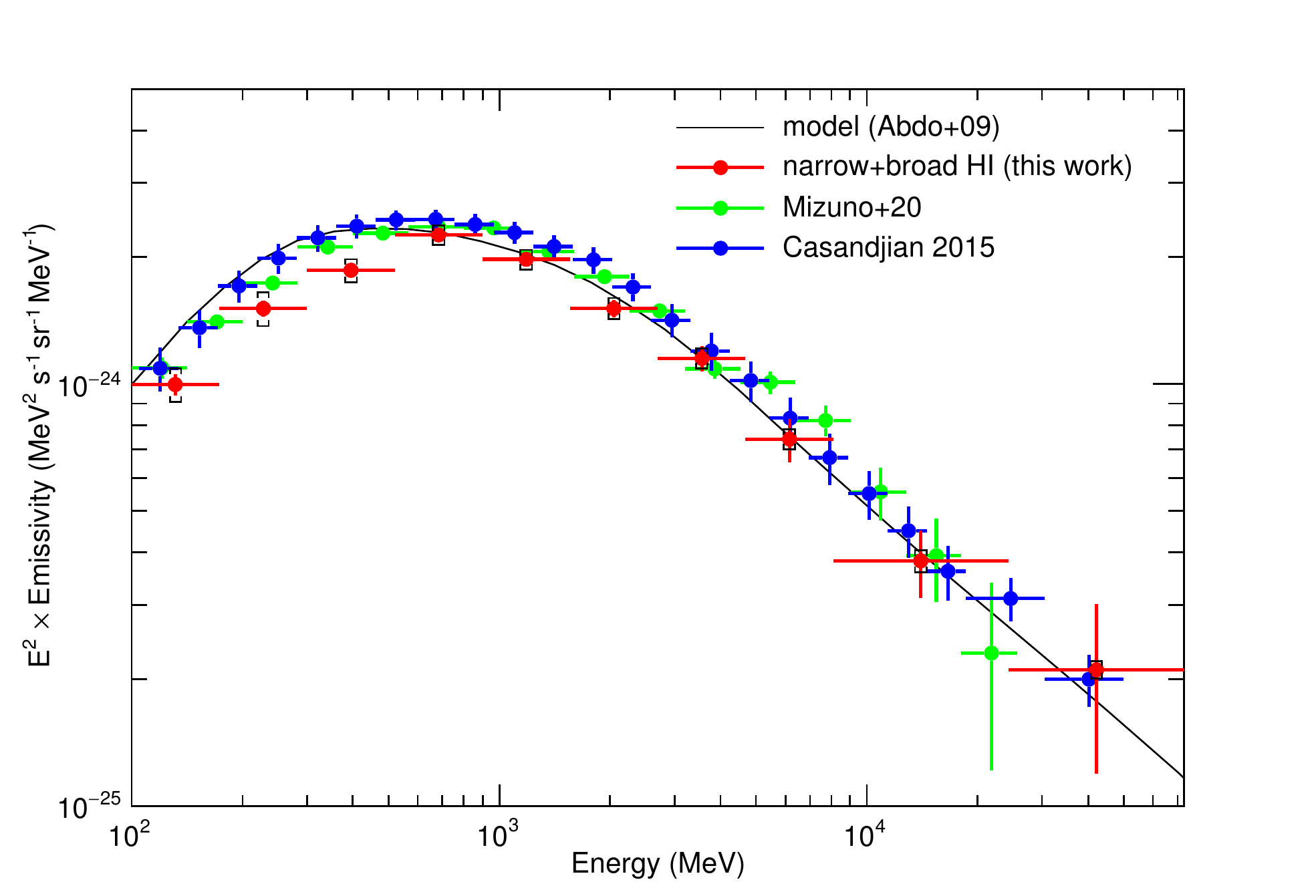}
{0.5\textwidth}{(a)}
\fig{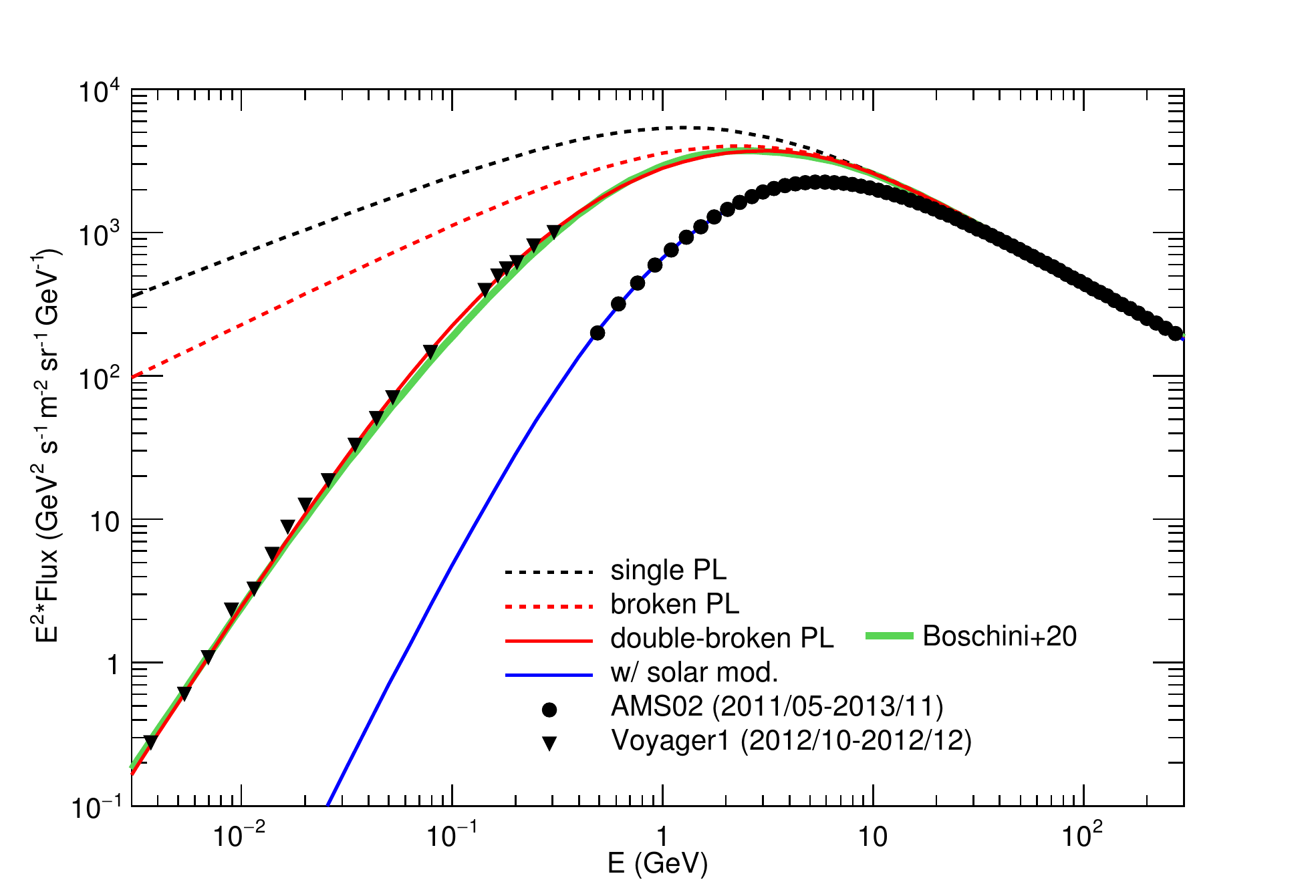}
{0.5\textwidth}{(b)}
}
\gridline{
\fig{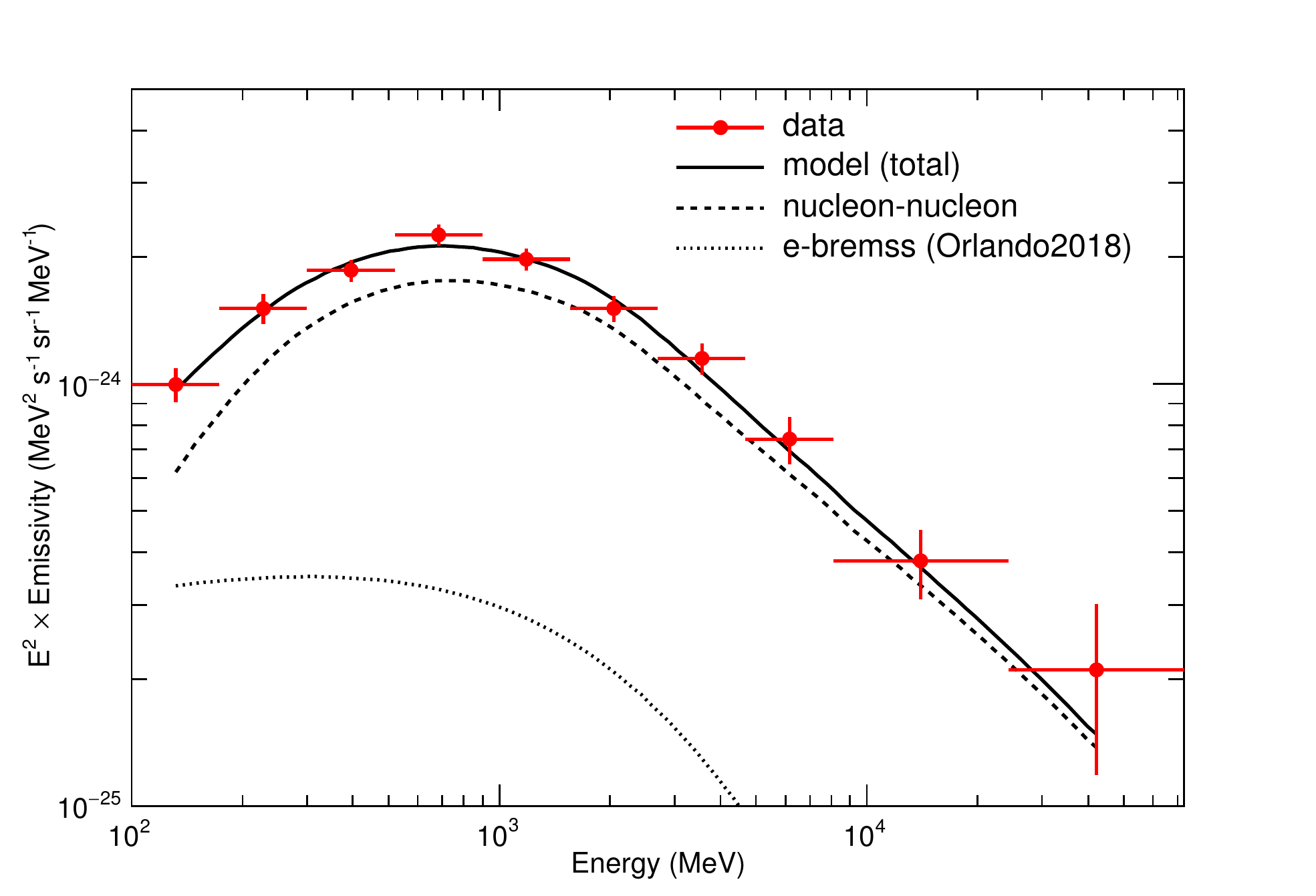}
{0.5\textwidth}{(c)}
\fig{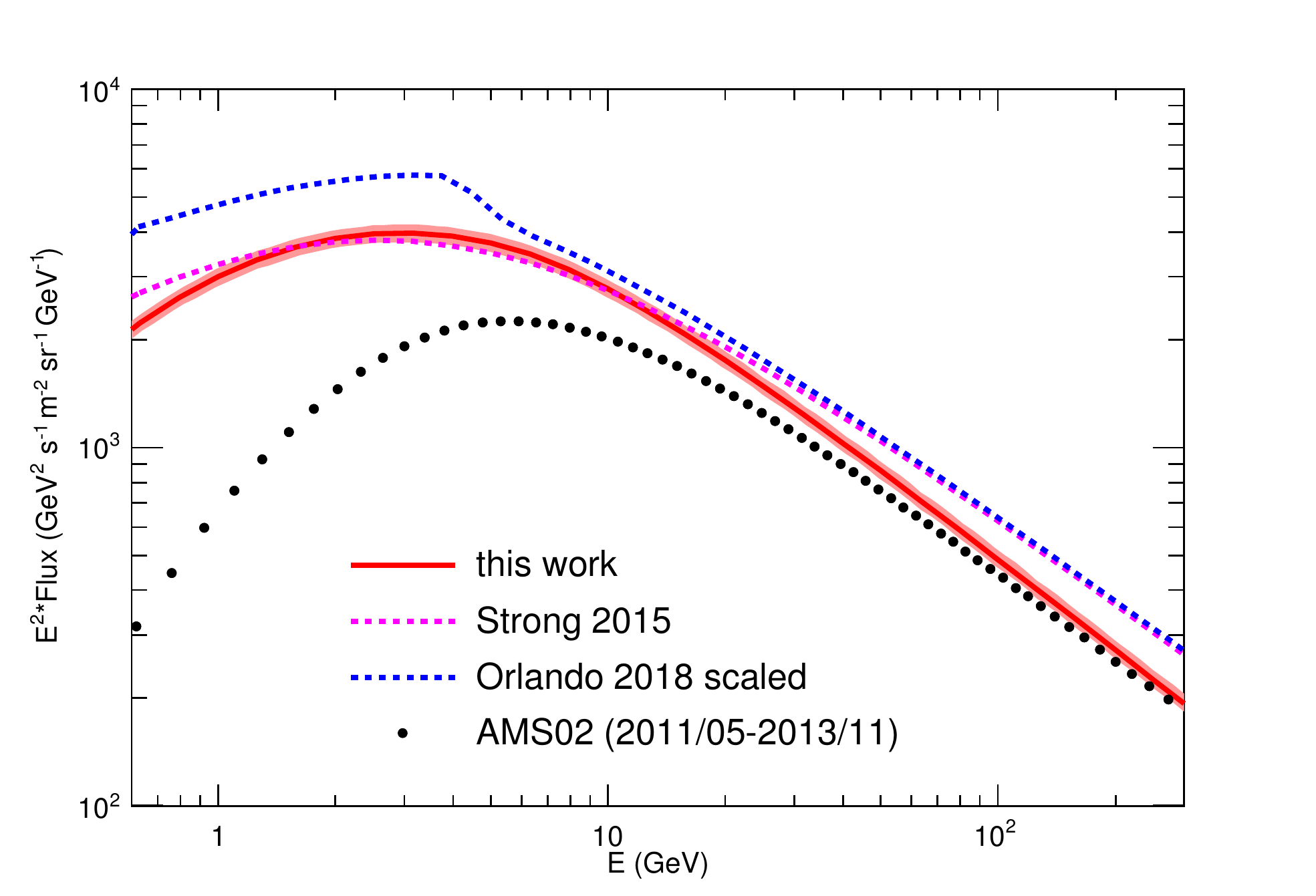}
{0.5\textwidth}{(d)}
}
\caption{
(a) The $\HI$ emissivity spectrum obtained in this study (red points)
compared with the model curve and the results
of relevant studies from \textit{Fermi}-LAT \citep{FermiHI2, Mizuno2020}.
Square brackets represent the systematic errors. 
The emissivity model adopted in this analysis is also shown as a solid black line.
(b) The CR proton LIS obtained by fitting the CR and $\gamma$-ray data. The LIS and the spectrum at the Earth
are shown by the solid red and blue lines, respectively. For reference, a spectrum without break and one
with only the high-energy break are shown by the dotted black and red lines, respectively.
The spectrum obtained by \citet{Boschini2020} is shown by the green line for comparison.
(c) The emissivity model spectrum obtained by fitting the CR and $\gamma$-ray data. 
The data is shown with the statistical and systematic errors summed in quadrature.
Contributions of hadronic interation
and electron bremsstrahlung are indicated. See Appendix~C for details of the model calculations.
(d) A comparison of LIS models based on $\gamma$-ray emissivities
(this work, that of \citet{Strong2015}, and that of \citet{Orlando2018})
in GeV energy range. 
The AMS-02 data is also shown.
Our LIS is multiplied by 1.07 to include the scaling factor for $\gamma$-rays.
The LIS model of \citet{Strong2015} has been obtained by fitting an analytical formula to the emissivity by \citet{FermiHI2},
and that of \citet{Orlando2018} has been obtained by scaling a CR propagation model called PDDE 
to fit the same emissivity spectrum.
The uncertainty of our LIS is shown by a light-red shaded band. 
Those for \citet{Strong2015} and \citet{Orlando2018} are not presented for clarity.
\label{fig:f8}
}
\end{figure}

\clearpage

\section{Summary}
We carried out a detailed study of the ISM and CRs using the \textit{Fermi}-LAT data in
the 0.3-72.9~GeV energy range 
at Galactic longitudes from $60\arcdeg$ to $120\arcdeg$ and Galactic latitudes from 
$-60\arcdeg$ to $-28\arcdeg$.
This region encompasses the MBM 53-55 clouds and the Pegusas loop.
We improved the ISM gas modeling over our previous work 
by using the Gaussian decomposition of the 21-cm line emission. 
We succeeded in distinguishing optically thin $\HI$, optically thick $\HI$, and CO-dark $\Htwo$ gas phases.
We found the fractions of $\HI$ optical depth correction
and of CO-dark $\Htwo$ to be
nearly equal, and we found the fraction of dark gas to be about 20\% of the total gas column density.
The CO-dark $\Htwo$ fraction is higher in the Pegasus loop than in the MBM~53-55 clouds, likely
due to the CO photodissociation.
While the $\HI$ emissivity spectrum agrees with the adopted model of the LIS above 1~GeV, there is a small deviation below 
a few hundreds of MeV. We fitted the CR spectra measured at/near the Earth and the measured $\gamma$-ray spectrum simultaneously,
and we obtained a spectral break in the proton LIS at ${\sim}7~\mathrm{GeV}$.
Our new emissivity spectrum relaxes the tension with the CR spectra directly measured at the Earth,
and agrees with the AMS-02 spectrum within 10\%.

\begin{acknowledgements}
The \textit{Fermi} LAT Collaboration acknowledges generous ongoing support
from a number of agencies and institutes that have supported both the
development and the operation of the LAT as well as scientific data analysis.
These include the National Aeronautics and Space Administration and the
Department of Energy in the United States, the Commissariat \`a l'Energie Atomique
and the Centre National de la Recherche Scientifique / Institut National de Physique
Nucl\'eaire et de Physique des Particules in France, the Agenzia Spaziale Italiana
and the Istituto Nazionale di Fisica Nucleare in Italy, the Ministry of Education,
Culture, Sports, Science and Technology (MEXT) and High Energy Accelerator Research
Organization (KEK) 
in Japan, and
the K.~A.~Wallenberg Foundation, the Swedish Research Council and the
Swedish National Space Board in Sweden.
 
Additional support for science analysis during the operations phase is gratefully
acknowledged from the Istituto Nazionale di Astrofisica in Italy and the Centre
National d'\'Etudes Spatiales in France. This work performed in part under DOE
Contract DE-AC02-76SF00515.

This work was partially supported by JSPS Grants-in-Aid for Scientific Research (KAKENHI) grant No.
17H02866 (T.M.) and by Core of Research for the Energetic Universe at Hiroshima University.
E.O. acknowledges the ASI-INAF agreement n. 2017-14-H.0 and the NASA Grant No. 80NSSC22K0495
GALPROP development is partially supported through NASA grant NNX17AB48G.
\end{acknowledgements}

\software{
{emcee \citep{emcee}},
{Fermitools (v2.0.0; Fermi Science Support Development Team 2019)},
{AAfrag \citep{AAfrag}}
}

\appendix

\section{Treatment of the Infrared Sources}
In the \textit{Planck} dust-model maps from Data Release 1, we identified several regions with high $T_{\rm d}$,
indicating localized heating by stars.
We refilled these areas in the $R$, $\tau_{353}$, and $T_{\rm d}$ maps
with the average of the peripheral pixels:
values in a circular region of radius $r_{1}$ are filled with the average
of the pixels in an annulus with inner radius $r_{1}$ and outer radius $r_{2}$.
For each region, the central position
($l, b$), $r_{1}$, and $r_{2}$ are summarized in Table~5.
Since the area of high $T_{\rm d}$ located near 3C~454.3 is large, we used a larger radius for it.
We found that the \textit{Planck} Data Release 2 maps are less affected by
infrared sources, and we had to mask only RAFGL~3068.

\floattable
\begin{deluxetable}{ccccc}[htb!]
\tablecaption{
Infrared sources excised and interpolated across
in the \textit{Planck} dust maps}
\tablecolumns{5}
\tablewidth{0pt}
\tablehead{
\multicolumn{2}{c}{Position} & \colhead{$r_{1}$} & \colhead{$r_{2}$} &\colhead{Object name} \\
\cline{1-2}
\colhead{$l$ (deg)} & \colhead{$b$ (deg)} & \colhead{(deg)} & \colhead{(deg)} &\colhead{}
}
\startdata
79.61 & $-30.25$ & 0.12 & 0.15 & \\
82.85 & $-50.65$ & 0.12 & 0.15 & \\
83.10 & $-45.46$ & 0.12 & 0.15 & \\
86.30 & $-38.20$ & 0.60 & 0.65 & 3C~454.3 (Active galactic nucleus)\\
87.46 & $-29.73$ & 0.12 & 0.15 & NGC~7339 (Radio galaxy)\\
87.57 & $-39.12$ & 0.12 & 0.15 & \\
93.53 & $-40.35$ & 0.12 & 0.15 & RAFGL~3068 (Variable star)\\
93.91 & $-40.47$ & 0.12 & 0.15 & NGC~7625 (Interacting galaxies)\\
97.29 & $-32.52$ & 0.12 & 0.15 & IC~5298 (Seyfert2 galaxy)\\
98.88 & $-36.55$ & 0.12 & 0.15 & NGC~7678 (Active galactic nucleus)\\
104.26 & $-40.58$ & 0.12 & 0.15 & \\
104.46 & $-40.14$ & 0.12 & 0.15 & \\
111.37 & $-36.00$ & 0.12 & 0.15 & \\
\enddata
\end{deluxetable}


\section{IC model}
We employed recent work by \citet{Porter2017} to construct the IC model template. 
They employed 3D spatial models for the CR source distribution and the ISRF.
They considered three different spatial distributions for the CR sources
(differentiated by the ratio of the smooth-disk component to the spiral-arm component) and three ISRFs.
The three CR source distributions are labeled SA0, SA50, and SA100;
SA0 corresponds to a 100\% (2D) disk, and SA100 corresponds to a 100\% spiral-arm contribution.
For the ISRF, they used a standard 2D ISRF (labeled Std) and two 3D ISRFs (labeled R12 and F98).
We tested all nine IC models
and a model used by \citet{Mizuno2016} (labeled 54\_77Xvarh7S) against the $\gamma$-ray data using our baseline gas model.
We found that the SA0 models give a better fit than the others in terms of log-likelihoods,
and that the difference among the three ISRF is minor.
We therefore decided to use the SA0-Std model in this study.

\section{CR and Gamma-Ray Fitting Framework}

To investigate the interstellar CR spectrum in detail, we have developed a framework that simultaneously fits
the CR and $\gamma$-ray data.
It models the LIS and the solar-modulation effect using analytical formulae,
and it uses a Markov chain Monte Carlo (MCMC) technique to constrain the model parameters.
Specifically, we use the emcee\footnote{
\url{https://emcee.readthedocs.io/en/stable/}} \citep{emcee}
python package that implements the affine-invariant ensemble sampler 
\citep[][]{Goodman2010}.
Solar modulation is taken into account using a force-field approximation \citep{Gleeson1968},
and the LIS is modeled as a power law of momentum with two breaks [see Equation~(6)].
In the $\gamma$-ray spectrum calculation, we take account of p-p, p-He, He-p, and He-He interactions individually
using the AAfrag package \citep{AAfrag} and parameterizations in \citet{Kamae2006}.
Specifically, we adopt the calculations using the AAfrag package above 10~GeV/n and use the parameterizations
by \citet{Kamae2006} below that energy.
For the p-p interaction, we use the non-diffractive component of \citet{Kamae2006},
since it smoothly connects to the AAfrag calculation.
For the p-He and He-p interactions, we use the p-p interaction model by \citet{Kamae2006} multiplied by a factor of 4 to 
connect it smoothly to the AAfrag calculation. For He-He interaction, we adopt a scale factor of 14.
The contribution of heavier nuclei to the $\gamma$-ray spectrum is small, and we use an
enhancement factor based on the formalism of \citet{Kachelriess2014}
to scale the flux up to account for all other elements in the CRs and the interstellar gas.
We adopt the spectra from \citet{Honda2004} and the abundances in the interstellar medium from \citet{Meyer1985}
for the heavier nuclei in the CRs and the ISM gas, respectively.
We also add an electron/positron bremsstrahlung model by \citet{Orlando2018}, specifically their
best propagation model called PDDE. In that work, constraints on the electron/positron LIS were obtained 
by fitting the CR direct measurements (Voyager~1 and AMS-02), the local synchrotron emission from radio to microwaves
(radio surveys and \textit{Planck} data), and the local $\gamma$-rays emissivity by \citet{FermiHI2}.
This method allows to obtain a consistent electron/positron LIS independent from assumptions on the solar modulation.

The framework reads the AMS-02 data (taken in 2011-2013), the Voyager~1 data (taken in 2012), and the \textit{Fermi}-LAT $\gamma$-ray data.
To disentangle a possible degeneracy between the shape of the LIS and solar modulation, we can use other CR data.
All the CR data is retrieved from the Cosmic-Ray Data Base \citep{Maurin2014},
but datapoints above 300~GeV are not used in the fitting.
The CR data and $\gamma$-ray data share the same spectral shape of the LIS, but the normalizations
(relative to that of AMS-02) are allowed to vary to account for possible systematic uncertainties.
The solar-modulation potential $\phi$ is also set free for each CR dataset, and it is set to be 0~V for Voyager~1 data.
The proton and He data of the same experiment and observational period share the common value of $\phi$.
Since the high-energy break 
is presumably due to a break in the interstellar diffusion coefficient,
CR protons and He ions share $\delta_{1}$ and a common value of rigidity for $p_\mathrm{br1}$.
$\alpha_{2}$ is also common among them.
We also limit parameter ranges as summarized in Table~7.
Except for Voyager~1, $\phi$ is limited within $\pm 15$\% of the value calculated based on \citet{Usoskin2017}
and the observational period.
For prior probabilities, we adopt the Gaussian distribution for LIS normalizations 
(relative to that of AMS-02); the standard deviations are 0.05 and 0.1 for CR data and $\gamma$-ray emissivity, respectively.
We adopt the flat distribution for other parameters.
The likelihood is calculated assuming the Gaussian distribution for each CR/$\gamma$-ray datapoints
(statistical error and systematic error are summed in quadrature).
The framework then runs the MCMC fitting to constrain the LIS parameters and $\phi$. 
The list of datasets used for this study and obtained values of $\phi$ and LIS normalization are
summarized in Table~8.

\floattable
\begin{deluxetable}{cc}[htb!]
\tablecaption{
Parameters ranges}
\tablecolumns{2}
\tablewidth{0pt}
\tablehead{
\colhead{parameter} & \colhead{range}
}
\startdata
proton normalization & 22.5-27.5 \\
$\alpha_{1}$ & 2.7--3.0 \\
$\alpha_{2}$ & 2.2--2.7 \\
$\alpha_{3}$ & $<0$ \\
$R_\mathrm{br1}$ (GV) & 2--10 \\
$R_\mathrm{br2}$ (GV) & 0.1--2 \\
$\delta_{1}$ & 0.05-2 \\
$\delta_{2}$ & 0.05-2 \\
$\phi$ & $(0.85-1.15) \times \phi_{0}$ \\
\enddata
\tablecomments{
Normalization is the flux [$\mathrm{c~s^{-1}~m^{-2}~sr^{-1}~(GeV/n)^{-1}}$] at $p = 10~\mathrm{GeV}$.
CR protons and He are assumed to share the value of the high-energy spectral break in rigidity.
Therefore breaks in rigidity ($R_\mathrm{br}$) are given instead of breaks in momentum.
$\phi_{0}$ is the reference value of $\phi$ calculated based on \citet{Usoskin2017} and the observational period.
}
\end{deluxetable}

\floattable
\begin{deluxetable}{ccc}[htb!]
\tablecaption{
Datasets and their $\phi$ and relative normalization}
\tablecolumns{3}
\tablewidth{0pt}
\tablehead{
\colhead{experiment} & \colhead{$\phi$} & \colhead{relative normalization} \\
\colhead{} & \colhead{(GV)} 
}
\startdata
AMS-02, 2011/05-2013/11 (proton, He) & $614.7 \pm 4.7$ & 1 \\
Voyager~1, 2012/10-2012/12 (proton, He) & 0 & 1 \\
AMS-01, 1998/06 (proton) & 472 \tablenotemark{a} & $1.028 \pm 0.009$ \\
BESS-PolarI, 2004/12 (proton, He) & $668.4 \pm 4.4$  & $1.012 \pm 0.004$ \\
BESS-PolarII, 2007/12-2008/01 (proton, He) & $386.5 \pm 3.7$  & $0.959 \pm 0.004$ \\
BESS-TeV, 2002/08 (proton, He) & $1055.0 \pm 7.5$  & $0.986 \pm 0.004$ \\
PAMELA, 2006/07 (proton) & $504.1 \pm 5.0$  & 1 \\
PAMELA, 2008/03-2008/04 (proton) & $404.6 \pm 4.8$ & 1 \\
PAMELA, 2010/01 (proton) & $293.7 \pm 5.4$  & 1 \\
$\gamma$-ray emissivity, this work & 0 & $1.066 \pm 0.025$ \\
\enddata
\tablecomments{
The value of $\phi$ is set to 0 for Voyager~1 data. 
LIS normalizations are scaled to that of AMS-02 and allowed to vary. 
Some experiments do not have enough high-energy datapoints to constrain the
normalization; in such a case, the relative normalization is fixed to 1.
}
\tablenotetext{a}{Although the best-fit value of $\phi$ is at the parameter limit, the spectrum is represented by the model well.}
\end{deluxetable}


\bibliography{sample631}{}
\bibliographystyle{aasjournal}



\end{document}